\documentstyle[12pt,epsf]{article}

\begin{document}                                 

% \nocite{*}                                 
% My default margin widths and so on unless overridden in the latex file
\setlength{\oddsidemargin}{0.25in}      % 1.25in left margin
\setlength{\evensidemargin}{0.25in}     % 1.25in left margin (even pages)
\setlength{\topmargin}{0.0in}           % 1in top margin
\setlength{\textwidth}{6.0in}           % 6.0in text - 1.25in rt margin
\setlength{\textheight}{9in}            % Body ht for 1in margins
\addtolength{\topmargin}{-\headheight}  % No header, so compensate
\addtolength{\topmargin}{-\headsep}     % for header height and separation
\setlength{\marginparwidth}{0.75in}                                 
% \setlength{\marginparsep}(0.05 in}                                 

% useful macros
\newcommand{\myprime}{^\prime}
\newcommand{\grad}{\nabla}
\newcommand{\mass}{{\cal M}}
\newcommand{\lapse}{N^{t}}
\newcommand{\shift}{N^{r}}

%\title{Back-Reaction and The Validity of The Semiclassical
%       Approach to Black Hole Evaporation}
\title{\begin{flushright}
\begin{small} PUPT-1584, hep-th/9512148 \end{small}
\end{flushright}
Back-Reaction and Complementarity in \\ 
       $1+1$ Dilaton Gravity}
\author{Vijay Balasubramanian\thanks{vijayb@puhep1.princeton.edu}~~and 
        Herman Verlinde\thanks{verlinde@puhep1.princeton.edu} \\  
        $^{*,\dagger}${\it Dept. of Physics, Princeton University,
Princeton,   NJ 08544} \\
           and \\
$^\dagger${\it Institute for Theoretical Physics, University of Amsterdam} }
\date{\today}                                 
\maketitle

\begin{abstract}
We study radiation from black holes in the effective theory produced
by integrating gravity and the dilaton out of $1+1$ dilaton gravity.
The semiclassical wavefunctions for the dressed particles show that
the self-interactions produce an unusual renormalization of the
frequencies of outgoing states.  Modes propagating in the dynamical
background of an incoming quantum state are seen to acquire large
scattering phases that nevertheless conspire, in the absence of
self-interactions, to preserve the thermality of the Hawking
radiation.  However, the in-out scattering matrix does not commute
with the self-interactions and this could lead to observable
corrections to the final state.  Finally, our calculations explicitly
display the limited validity of the semiclassical theory of Hawking
radiation and provide support for a formulation of black hole
complementarity.
\end{abstract}

\section{Introduction}
Much of the controversy surrounding discussion of the black hole
information loss problem hinges on the question of the validity of the
semiclassical approximation to gravity.  In Hawking's original
analysis it was argued that back-reaction effects would be small and
that the Hawking modes could be considered as propagating in a
background unaffected by their presence (\cite{hawking1},
\cite{hawking2}).  This assumption seems to lead inevitably to the
conclusion that black hole evaporation leads pure states to evolve
into mixed states.  In recent years there have been a growing number
of advocates for the point of view that the semiclassical theory
breaks down in such a way as to restore coherence of the final state
(\cite{thooft1}, \cite{stu}, \cite{sv21,sv22,kv2}).  Some of these
authors have claimed that within the context of a local field theory
the reasoning employed by Hawking is correct until late into the
evaporation and therefore exotic string theoretic effects are
necessary to resolve the information paradox (\cite{lowe}).  Others
have claimed that the breakdown occurs sufficiently early in the
lifetime of the black hole to be visible within the context of the
semiclassical theory itself (\cite{sv21,sv22,kv2},\cite{keski}).
Still others have argued that although the standard semiclassical
results do break down, the low energy theory remains under sufficient
control to permit explicit computation of nonthermal corrections to
the Hawking radiation (\cite{kw2,kw3}).

In this paper we will attempt to shed some light on these competing
viewpoints in the context of $1+1$ dilaton gravity coupled to massless
scalar matter.  Using the techniques of~\cite{fmp} and~\cite{kw2} 
we will completely integrate out the dilaton and the
graviton and work with the effective quantum mechanics of
gravitationally dressed particles.  As such we have presumably
completely included all the effects of gravitational and dilatonic
fluctuations in the underlying theory.  The resulting quantum
mechanics is too nonlinear for full canonical or path-integral
quantization.  Consequently we work in the WKB approximation and
develop the leading corrections to the wavefunctions of particles
propagating in a dilaton gravity background in the limit of small
$\hbar$.  These improved wavefunctions can then be used to repeat the
traditional Hawking analysis thereby providing the leading
back-reaction corrections to black hole radiation.

We work with the quantum mechanics of dressed particles as opposed to
a dressed field theory because it is significantly easier to obtain
the effective particle theory.  We are actually interested in
understanding the effects of back-reaction on the effective second
quantized system.  It is possible to recover some such insights from
the effective first quantized theory by identifying the second
quantized operators that produce the phenomena that we observe in a
basis of states in the first quantized language.  Our philosophy for
studying the back-reaction effects is the opposite of the approach
pursued in~\cite{cghs}.  Those authors essentially integrated out the
matter fields in order to include the back-reaction, while we
integrate out the graviton and the dilaton.

Having obtained the effective quantum mechanics, we examine two
situations in detail.  First of all, we study the effect of the
gravitational dressing of a single outgoing particle on the Hawking
flux.  The analogous dressing of spherically reduced Einstein gravity
has been found to give energy dependent shifts of the Hawking
temperature (\cite{kw1,kw2}).  However, we find the
temperature for radiation of dressed particles in $1+1$ dilaton
gravity remains unchanged.
%(This is not surprising because the Hawking
%temperature in this system depends only on the cosmological constant
%$\lambda$ and the gravitational self-interactions of matter particles
%are not expected to renormalize $\lambda$. )  
This is so despite the fact that the wavefunctions and associated
Bogliubov coefficients are significantly modified by the
self-interactions which produce an unusual renormalization of the
Kruskal frequencies of outgoing states.

  Next we examine the outgoing radiation in the dynamical background
of a single incoming quantum particle in an approximation where the
self-interactions are turned off.  The wavefunctions responsible for
the late time radiation acquire large scattering phases that depend on
the incoming state.
%In the case that the incoming particle is placed
%in a classical shockwave, we reproduce to leading order the shift
%interactions described in~\cite{thooft1} and~\cite{sv21,sv22,kv2}.
In general we are able to identify the effect of an arbitrary incoming
quantum wavefunction on the outgoing wavefunctions.  These effects
translate into phase shifts of the Bogliubov coeffcients that
determine the structure of the vacuum misalignment between the horizon
and infinity.  In the full second quantized treatment, these phase
shifts should be operator valued expressions connecting the incoming
and outgoing states.  Since the WKB approximation can be understood in
part as a replacement of operators by their expectation values, our
results allow us to identify the S matrix operator entangling
infalling and outgoing states, at least up to certain normal ordering
problems.  The leading terms suggested by~\cite{thooft1}
and~\cite{sv21,sv22,kv2}, as well as subleading corrections are
identified.  Despite the entanglement produced by this operator, the
radiation is seen to be thermal in the approximation in which the
self-interactions have been turned off.  However, we argue that the
non-commutativity of the in-out scattering operator with the
self-interactions could bring out some additional information about
the structure of the infalling state.

Finally, we examine the validity of the semiclassical calculations
that give these results.  We find that the computation of the
Bogliubov coefficients relating horizon states to asymptotic states is
not reliable in the presence of infalling matter.  The semiclassical
method is only reliable when the energies of the various particles
involved are small.  It is seen that even a small incoming energy
density leads to huge shifts in the energies of the outgoing states.
The size of these shifts grows exponentially in time and suggests a
rapid breakdown of semiclassical methods.  Moreover, we find that
there are two complementary, semiclassically controlled Hilbert spaces
with which the system can be described.  One is appropriate to
observers entering the black hole and another is is suitable for
observers of the Hawking radiation.  This lends support to the idea of
black hole complementarity.

\section{Classical Solutions of $1+1$ Dilaton Gravity}
\label{sec:classic}
In this paper we will be integrating the dilaton and the graviton out
of $1+1$ dilaton gravity coupled to a fixed number of matter particles.
Since the theory is two dimensional, the fields have no propagating
degrees of freedom and ``integrating out'' amounts to fixing a gauge
that is consistent with the constraints induced by the presence of
matter particles.   The family of gauges we will choose is quite
unusual and is closely related to the interesting parametrization of the
Schwarschild black hole adopted in~\cite{kw1}.   In order to have
intuitions for the effective theory it is useful to begin by
displaying the classical black hole solutions of dilaton gravity
in the same gauge so that we know how the space is sliced in the
absence of any back-reacting particles.   The action defining $1+1$
dilaton gravity is:
\begin{equation}
\label{eq:action}
S_D = \frac{1}{2\pi} \int d^2x \sqrt{-g} e^{-2\phi} \left[{\cal R} + 4
(\grad\phi)^2  + 4 \lambda^2\right]
\end{equation}
with $\phi$, ${\cal R}$ and $\lambda$ being the dilaton, Ricci scalar
and cosmological constant respectively.\footnote{This action can be
derived in a number of ways.  First of all, it is the low energy
effective action arising from $SL(2,R)/U(1)$ coset theory given
suitable conventions for the normalization of the dilaton and the
cosmological constant (\cite{witten}).  It is also the effective field
theory that operates within the throats of four dimensional,
near-extremal, magnetically charged, dilaton black holes
(\cite{cghs}). Finally, the spherical reduction of Einstein gravity
gives an action identical to that in Equation~\ref{eq:action} with the
cosmological constant moved out of the parentheses.}

\begin{figure}                                 
\begin{center}                                 
\leavevmode                                 
\epsfxsize=3in
\epsfysize=1.5in                                 
\epsffile{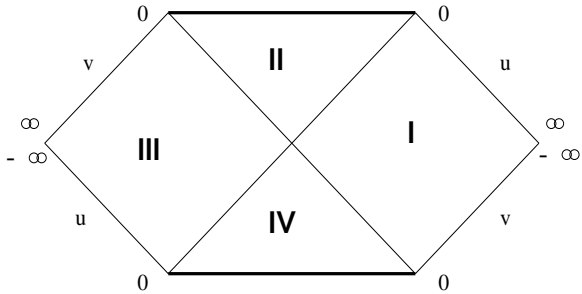}                                 
\end{center}                                 
\caption{Penrose Diagram For u-v Coordinate Patch \label{fig:Kruskal}}
\end{figure}                                 

\begin{figure}                                 
\begin{center}                                 
\leavevmode                                 
\epsfxsize=4in
\epsfysize=1.5in                                 
\epsffile{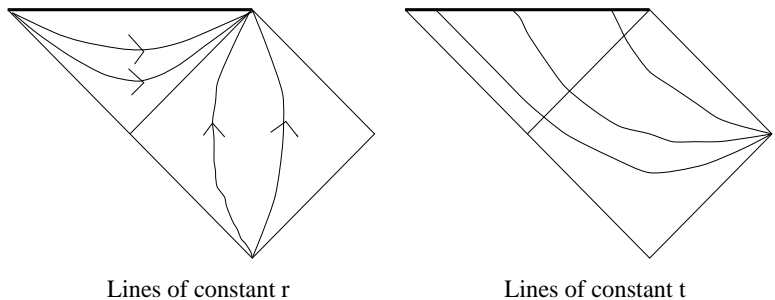}                                 
\end{center}                                 
\caption{Surfaces of Constant r and t \label{fig:rtconst}}
\end{figure}

Classical solutions of this action and several close variants have
been analyzed in detail in a number of papers.  (For
example, see~\cite{mandal}, \cite{cghs} and \cite{rst}.  For a good review
consult
\cite{thorlacius}.)  There is a spectrum of black hole solutions which
can be written in Kruskal-like coordinates as:
\begin{equation}
e^{-2\phi} = \frac{M\pi}{\lambda^2} - \lambda^2 uv \: \: ; \: \:
ds^2 =  \frac{-du\,dv}{(M\pi)/\lambda^2 - \lambda^2 uv} 
\end{equation}
where $M/\lambda$ can be shown to be the mass of the black hole. (The
factor of $\pi$ which is absent in the solution presented by
\cite{thorlacius} arises from a difference in conventions.)  A Penrose
diagram of this solution is displayed in Figure~\ref{fig:Kruskal}. Now
consider the following coordinate transformation:
\begin{equation}
u = \frac{1}{\lambda} e^{-\lambda t} \left(e^{\lambda r} -
\frac{\sqrt{M\pi}}{\lambda} \right) \: \: ; \: \:  
v = - \frac{1}{\lambda} e^{\lambda t} \left(e^{\lambda r} +
\frac{\sqrt{M\pi}}{\lambda} \right) 
\end{equation}
The metric, and the dilaton are now given by:
\begin{equation}
\label{eq:newcoords}
e^{-2\phi} = e^{2\lambda r} \: \: ; \: \: ds^2 = -dt^2 + (dr +
\frac{\sqrt{M\pi}}{e^{\lambda r} \lambda} dt)^2 \label{eq:met1}
\end{equation}
The important point about these coordinates is that the metric is
stationary, and regular at the horizon while at the same time being
asymptotically flat.  Consequently, the conserved energy defined as
the generator of translations with respect to $t$ is the Hamiltonian
of the system and even applies to states in the interior of the black
hole.  The new $r$, $t$ coordinates give a patch covering regions $I$
and $II$ of Figure~\ref{fig:Kruskal}. By changing the sign of $t$ in
the metric~\ref{eq:met1} we patch regions $I$ and $IV$.  By further
judicious changes of sign in the relations defining $u$ nd $v$ in
terms of $r$ and $t$ we get patches covering regions $II$ and $III$
and $IV$ and $III$. Figure~\ref{fig:rtconst} displays the surfaces
of constant $r$ and $t$ in the metric (\ref{eq:met1}) for regions $I$ and
$II$ that are of interest to us.  In this parametrization the horizon
is found at $\exp{-\phi} = \exp{\lambda r} = \sqrt{M\pi}/\lambda$ and
the singularity is at $\exp{-\phi} = 0$.

With these classical solutions in hand we can repeat Hawking's
calculation to evaluate the radiation streaming out of the black hole.
The radiation arises because of a misalignment of the vacua defined
with respect to inertial observers at the horizon and at infinity.
The vacuum that is defined as the state annihilated by all states of
positive Kruskal frequency is found to contain a thermal spectrum
when probed near future infinity.  On the other hand the vacuum
defined as the state annihilated by the asymptotic modes appears
singular to inertial observers at the horizon.  Since the horizon
should not be a locally distinguished location, we conclude that the
physical vacuum is annihilated by the Kruskal modes and that,
consequently, there is radiation at infinity. In this paper we will
improve upon the classic calculations by including the self-interaction
of the outgoing radiation in the presence of incoming matter.  For
purposes of comparison it is instructive to first reproduce the
Hawking calculation in the context of $1+1$ dilaton gravity coupled to a
massless scalar field.  (See Giddings and Nelson for a clear
discussion of a related scenario (\cite{gn}).)  In order to do this we
must solve the massless wave equation in the background~(\ref{eq:met1})
to find a complete set of energy eigenstates and a complete set of
Kruskal momentum eigenstates.  The former define the particle spectrum
measured by the asymptotic observer and the latter are the states that
have definite frequency with respect to inertial observers at the
horizon.  Using the conformal flatness of the metric it is easy to
show that the energy eigenstates are:
\begin{equation}
\phi_\pm = \frac{1}{\sqrt{2\omega}} \exp\left\{-i\omega t \pm i
(\omega/\lambda)\ln{(e^{\lambda r} \mp \sqrt{M\pi}/\lambda)}\right\}
\end{equation}
where the upper sign refers to outgoing waves and the lower sign
describes infalling waves. Note that the outgoing energy eigenstates are
singular at the horizon and therefore cannot be extended into region
$II$.  The Kruskal eigenstates that define the particle spectrum at
the horizon are found to be:
\begin{eqnarray}
\psi_+ &=& e^{i\omega u} = \exp\left\{\frac{i\omega}{\lambda}
e^{-\lambda t} (e^{\lambda r} - \frac{\sqrt{M\pi}}{\lambda})\right\}
 \label{eq:kruskalout} \\
\psi_- &=& e^{i\omega v} = \exp\left\{\frac{-i\omega}{\lambda}
e^{\lambda t} (e^{\lambda r} + \frac{\sqrt{M\pi}}{\lambda})\right\}
\label{eq:kruskalin}
\end{eqnarray}

In the region $I$ in Figure~\ref{fig:Kruskal} the outgoing part of a
massless scalar field can be expanded in either of the sets of modes
$\phi_+$ or $\psi_+$.  We write $\Phi_+ = \int d\omega \left[a_\omega
\psi_{+\omega} + a^\dagger_\omega \psi_{+\omega}^* \right] = \int
d\omega \left[b_\omega \phi_{+\omega} + b^\dagger_\omega
\phi_{+\omega}^* \right]$.  Since each of the sets $\phi_{+\omega}$
and $\psi_{+\omega}$ is complete we can write:
\begin{eqnarray}
\phi_{+\omega} &=& \int_0^\infty d\omega\myprime \left[
\alpha_{\omega\omega\myprime} \psi_{+\omega\myprime} +
 \beta_{\omega\omega\myprime} \psi_{+\omega\myprime}^* \right] \nonumber \\
\psi_{+\omega\myprime} &=& \int_0^\infty d\omega\myprime \left[
\alpha_{\omega\omega\myprime}^* \phi_{+\omega} - 
 \beta_{\omega\omega\myprime} \phi_{+\omega}^* \right] \label{eq:bog1}
\end{eqnarray}
where the second equation follows from the first in view of the fact
that the sets of states $\{\phi_{+\omega}\}$ and $\{\psi_{+\omega}\}$ are
orthonormal under a Klein-Gordon inner product.  Analogous relations
can be computed between the creation and annihilation operators
associated with $\psi_+$ and $\phi_+$.  Note that the coefficients
$\beta$ measure the mixing between positive and negative frequencies
that is responsible for the Hawking flux.  From the second of the
Equations~\ref{eq:bog1} we see that the Bogliubov coefficients
$\alpha$ and $\beta$ can be computed by projecting out the components
of $\psi$ that have definite frequency with respect to $t$.
\begin{eqnarray}
\alpha_{\omega\omega\myprime}^* &=& \frac{1}{\phi_{+\omega}(r)} \int dt
e^{i\omega t} \psi_{+\omega\myprime}   \nonumber \\
\beta_{\omega\omega\myprime}^* &=& - \frac{1}{\phi_{+\omega}(r)} \int dt
e^{i\omega t} \psi_{+\omega\myprime}^* \label{eq:bogints}
\end{eqnarray}
The $\phi_{+\omega}(r)$ in the denominator is the spatial part of the
energy eigenstate $\phi_{+\omega}$.\footnote{The Bogliubov
coefficients may also be evaluated by taking inner products between
the two bases on spatial surfaces.  However, in the presence of
backreacting particles it is not clear how to define the necessary
spatial surfaces and inner products.  In fact it is much more physical
to formulate the Bogliubov tranformation as a Fourier transform with
respect to time that projects out the components of definite frequency
in the Kruskal eigenstates.}  These integrals can be computed exactly
and give:
\begin{eqnarray}
\alpha_{\omega\omega\myprime}^* &=&
\sqrt{\frac{\omega}{|\omega\myprime|}}
\left(\frac{\omega\myprime}{i\lambda}\right)^{i\omega/\lambda} 
\frac{\Gamma(-i\omega/\lambda)}{\lambda}  
\nonumber \\
\beta_{\omega\omega\myprime}^* &=&
- \alpha_{\omega\omega\myprime}^*
e^{-\pi\omega/\lambda}
\end{eqnarray}
where $\beta$ can be simply computed from $\alpha$ by analytically
continuing $\omega\myprime \rightarrow - \omega\myprime$.  It can be
shown that since $\alpha_{\omega\omega\myprime} /
\beta_{\omega\omega\myprime}$ is independent of $\omega\myprime$, the
black hole radiates a flux $F(\omega)$ given by:
\begin{equation}
F(\omega) =
\frac{1}{\left|\frac{\alpha_{\omega\omega\myprime}}{\beta_{\omega\omega\myprime}}
\right|^2 - 1}= \frac{1}{e^{\frac{2\pi\omega}{\lambda}} - 1}
\end{equation}
In other words, the flux is thermal with a temperature of
$2\pi/\lambda$.  (We can see this also from the fact that Kruskal
eigenstates are clearly periodic in imaginary time with period
$2\pi/\lambda$.)  All correlation functions in the final state can
also be shown to be precisely thermal at late times (\cite{gn,bd}).

In the above analysis we made no mention of the states in the interior
of the black hole.  As discussed in~\cite{gn} we must augment the
$\phi$ basis by adding modes defined in region $II$.  Since the states
in the interior of the black hole are inaccessible to asymptotic
observers, we must trace over these interior modes to describe
experiments localized in region $I$.  This yields a density matrix
that is purely thermal. If this thermality persists all the way
through to the endpoint of black hole evaporation, a pure inital state
of the world will have evolved into a mixed state.  This is the
content of the information loss problem.  In the next section we will
construct the effective theory of particles propagating in a dilaton
gravity background and ask whether the self-interactions and
interactions with infalling matter modify the conclusions regarding
thermality of the radiation.

It is important to note that in order to compute the misalignment of
the Kruskal and asymptotic vacua it is necessary to understand the
behaviour of all the excited states since the Bogliubov coefficients
depend on them.  We are going to find that there are large effects
that threaten the validity of semiclassical computations but that
these effects vanish in the Kruskal vacuum state.  For this reason it
may appear that the effects we compute are not important because, as
discussed above, we specify the initial state to be the Kruskal
vacuum.  However, the computation of the misalignment between the
Kruskal and asymptotic vacua requires knowledges of the behaviour of
the excited states at the horizon.  We will see that these excited
states are very sensitive to self-interaction as well as to
interaction with infalling states.

\section{Dressed Particles in Dilaton Gravity}
\label{sec:dress}
In this section we will construct an effective theory of matter
particles by integrating the graviton and dilaton out of the action
for $N$ point particles coupled to $1+1$ dilaton gravity.  We will 
work in the Hamiltonian formulation of the theory for which purpose it
is useful to introduce the ADM parametrization of the metric:
\begin{equation}
\label{eq:ADM}
ds^2 = - {\lapse}^2 \,dt^2 + L^2(dr + \shift \, dt)^2
\end{equation}
where $\lapse$, $L$ and $\shift$ are functions of the coordinates $r$ and $t$.
We also define $R \equiv \exp{-\phi}$ in terms of which the Lagrangian in
Equation~\ref{eq:action} becomes $L = R^2 \, {\cal R} + 4 (\grad R)^2 + 4
R^2 \, \lambda^2$.   In this parametrization, the Hamiltonian formulation
of $1+1$ dilaton gravity can be shown to be:
\begin{equation}
\label{eq:ham}
S_D = \int dr dt \left[\pi_l \dot{L} + \pi_R \dot{R} 
		- \lapse H_t - \shift H_r \right] - \int dt M_{ADM} 
\end{equation}
where $\pi_L$ and $\pi_R$ are the canonical momenta of the fields $L$
and $R$, $M_{ADM}$ is the ADM mass of the system, and $H_t$ and $H_r$
are given below: 
\begin{eqnarray}
H_t &=& -\frac{1}{\pi} \left[ 2LR^2\lambda^2 + \frac{2(R\myprime)^2}{L} 
                  - \left( \frac{(R^2)\myprime}{L} \right)\myprime
		\right] + \frac{\pi}{2R^2} \left[L \pi_L^2 - \pi_L
                 \pi_R R \right]    \nonumber \\
H_r &=& \pi_R R\myprime - \pi_L\myprime L \label{eq:rconsrt}
\end{eqnarray}
We now couple $n$ point particles to the system.   In Hamiltonian
form, the matter part of the action is:
\begin{eqnarray}
S_M = \sum_{i=i}^n \left\{ \int dt p_i \dot{r_i} - \int dt\,dr\,
(\lapse H_{tM}(r_i) + \shift H_{rM}(r_i) ) \right\} \nonumber \\
H_{tm}(r_i) = \sqrt{(p_i/L)^2 + m^2}\, \delta(r - r_i(t)) \: \: ; \: \:
H_{rm}(r_i) = -p_i \,  \delta(r - r_i(t))
\end{eqnarray}
where $m$ is the mass of the particle which we will take to be zero.
Adding $S_M$ and $S_D$ describes dilaton gravity coupled to $n$ point
particles with $M_{ADM}$, the total energy of the system, specifying
the Hamiltonian.  Note that in the combined action there are no time
derivatives of $\lapse$ and $\shift$.  Consequently, these quantities
can be integrated out of the action generating the constraints that
$H_t + H_{tM}$ and $H_r + H_{rM}$ must vanish.

The elimination of the metric and dilaton from this action following
the methods of~\cite{fmp} and~\cite{kw2} is described in
Appendix~\ref{sec:elim}.  Since the technical details are quite
confusing we will describe the basic idea of the construction here.
Since the dilaton and the graviton are not dynamical in two dimensions
we can eliminate them by a choice of gauge.  However, the gauge must
be consistent with constraints that arise from varying $\lapse$ and
$\shift$.  Put another way, the $n$ matter particles cause kinks in
the geometry that move around with the particles.  The action for the
motion of these kinks is the only non-trivial contribution of the
geometry to the dynamics.  In eliminating the metric and the dilaton
from the Lagrangian we have to be careful to dress the particles with
the action of the kinks in the geometry that they generate.  The basic
idea of~\cite{fmp} and~\cite{kw2} is to compute the contribution of a
kink in the geometry by integrating up the constrained action for the
geometry in the presence of a particle and then differentiating with
respect to time to recover the constrained Lagrangian.  At first sight
this seems a rather strange procedure - one may wonder why we could
not directly fix a gauge consistent with the constraint.  The reason
for pursuing this awkward procedure is that the constraint produced by
the propagation of a point particle contains delta function
singularities and by first integrating past the singularities and then
differentiating back we ensure that we are not missing any
contributions to the dressed action.

\begin{figure}                                 
\begin{center}                                 
\leavevmode                                 
\epsfxsize=5in
\epsfysize=1.5in                                 
\epsffile{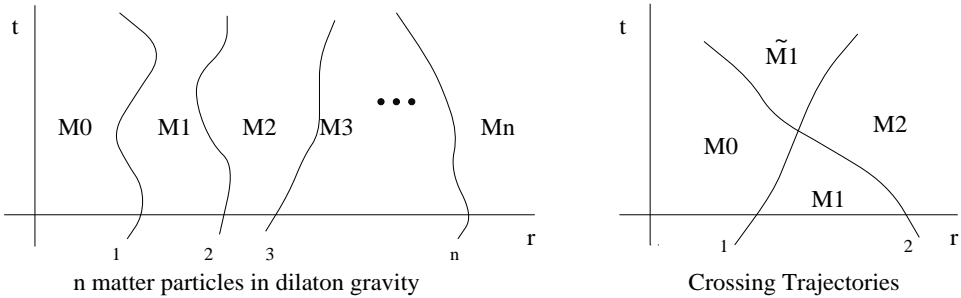}                             
\end{center}                                 
\caption{Particles Propagating in Dilaton Gravity \label{fig:particles}}
\end{figure}                                 

\subsection{Trajectories That Do Not Intersect}
\label{sec:noninter}
In Appendix~\ref{sec:elim} we have implemented the procedure for
eliminating the dilaton and the graviton described in the previous
section. It turns out that the geometry between the particles looks
like a slice of a black hole of fixed mass.  This is illustrated in
Figure~\ref{fig:particles} where $M_i/\lambda$ labels the mass outside
the ith particle and $M_0/\lambda$ is the mass of the background black
hole, if any.  Furthermore, in a gauge with $R= \exp{\lambda r}$ and
$L = 1$ we obtain the following effective lagrangian for dressed
massless particles:
\begin{eqnarray}
\label{eq:lag}
L = \left(\frac{2}{\pi}\right) \sum_{i=1}^{N}
 &\dot{r_i}& \left\{ e^{\lambda r_i} \left( \sqrt{M_{i-1}\pi} -
\sqrt{M_i \pi} \right)   \right. \nonumber \\
&\mbox{}& \left.  
- \eta_i \lambda e^{2\lambda r_i}
\ln\left(\frac{e^{\lambda r_i} - \eta_i \sqrt{M_i
\pi}/\lambda}{e^{\lambda r_i} - \eta_i \sqrt{M_{i-1} \pi}/\lambda}
\right) \right\} \: - \: \frac{M_n - M_0}{\lambda} 
\end{eqnarray}
where $\eta_i$ is $+1$ for outgoing particles and $-1$ for infalling
particles.  This expression should be read as a Hamiltonian
formulation of the Lagrangian ($L = \sum_i p_i \dot{r_i} - H$) so that
$(M_n - M_0)/\lambda$ is identified as the Hamiltonian of the system
and the canonical momentum of the ith particle is:
\begin{equation}
\label{eq:can}
p_{ci}= e^{\lambda r_i} \frac{2}{\pi}\left( \sqrt{M_{i-1}\pi} -
\sqrt{M_i \pi} \right) - \eta_i \lambda e^{2\lambda r_i}
\frac{2}{\pi}
\ln\left(\frac{e^{\lambda r_i} - \eta_i \sqrt{M_i
\pi}/\lambda}{e^{\lambda r_i} - \eta_i \sqrt{M_{i-1} \pi}/\lambda}
\right)
\end{equation}
This expression for the canonical momentum implicitly defines $M_i$ in
terms of  $M_{i-1}$ and $r_i$ and this chain of relations defines
$M_n/\lambda$, the Hamiltonian, in terms of the coordinates and momenta
of each of the particles.   $M_i/\lambda$ should  be thought of as the mass
parameter of the geometry outside the ith particle and so $(M_i -
M_{i-1})/\lambda $ can be understood as the energy of the ith particle.   

\begin{figure}                                 
\begin{center}                                 
\leavevmode                                 
\epsfxsize=4in
\epsfysize=1.5in                                 
\epsffile{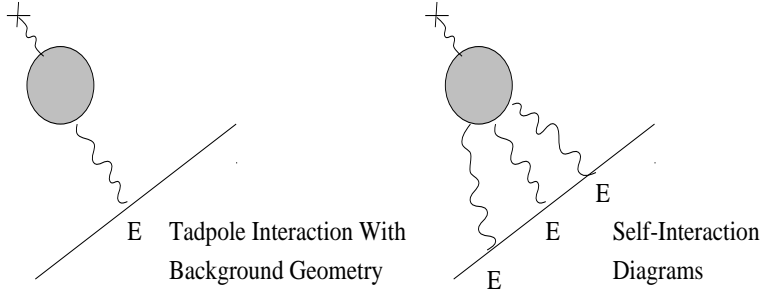}                                 
\end{center}                                 
\caption{Diagrams Contributing to The Effective Lagrangian \label{fig:diags}}
\end{figure}                                 

\subsection{Meaning Of The Effective Lagrangian}
In order to understand the meaning of this effective Lagrangian it is
useful to consider a single dressed particle whose Lagrangian is given
by:
\begin{eqnarray}
\label{eq:1part}
L = \left(\frac{2}{\pi}\right)
 &\dot{r}& \left\{ e^{\lambda r} \left( \sqrt{M_{0}\pi} -
\sqrt{M_1 \pi} \right) \right .\nonumber \\
&\mbox{}&
\left. - \eta \lambda e^{2\lambda r}
\ln\left(\frac{e^{\lambda r} - \eta \sqrt{M_1
\pi}/\lambda}{e^{\lambda r} - \eta \sqrt{M_{0} \pi}/\lambda}
\right) \right\} \: - \: \frac{M_1 - M_0}{\lambda} 
\end{eqnarray}
Here $M_0$ is the mass of the background black hole and $M_1$ is the
ADM mass of the geometry.  At large $r$ we can expand in powers of
$\exp{-\lambda r}$ and we find that $p_c = \eta (M_1 - M_0)/\lambda +
O(\exp{-\lambda r})$.  This correctly tells us that near infinity, to
leading order, the momentum of the massless particle equals its
energy.  We wish to interpret the effective Lagrangian in terms of the
diagrams that have contributed to the dressing of the particle.  In
flat space we should have $p = E$, and in curved space the interaction
with the background gives the classical equation $p= E f(r)$ where
$f(r)$ is a function of $r$ that can be read off from the geometry.
In diagrammatic terms, the diagram proportional to $E$ in
Figure~\ref{fig:diags} gives the classical description of a massless
particle in curved space and the diagrams with more graviton legs
represent the self-interaction corrections.  Indeed, writing $E = (M_1
- M_0)/\lambda$ and linearizing Equation~\ref{eq:1part} we find that:
\begin{equation}
p = \eta E \frac{e^{\lambda r}}{e^{\lambda r} - \sqrt{M_0 \pi}/\lambda}
\end{equation}
This is exactly the relation between momentum and energy for a
classical massless particle in the dilaton black hole in
Equation~\ref{eq:newcoords}.  This agreement increases our confidence
that we have correctly derived the effective theory and suggests that
an expansion of $p_c$ in powers of $E = (M_1 - M_0)/\lambda$ amounts
to a summation of the n-graviton self-interaction diagrams in
Figure~\ref{fig:diags}.  The Lagrangian in Equation~\ref{eq:lag}
contains graviton exchanges between all the particles in the system.
We can truncate to exchanges of $k$ gravitons by expanding to the
appropriate powers in the energies of each particle.  Note that for
outgoing particles ($\eta = +1$), Equation~\ref{eq:1part} tells us
that a particle of finite energy has a momentum that blows up as
$\exp{\lambda r} \rightarrow \sqrt{M_1 \pi}/\lambda$.  From the
equations of motion computed below we will see that $\sqrt{M_1
\pi}/\lambda$ is the location of the horizon and the blowup is
symptomatic of the huge redshifts between the horizon and infinity.

\subsection{Crossing Trajectories}
Having understood the dressed Lagrangian describing particle
trajectories that do not intersect we turn to the description of
particles that cross each other.  In the right hand
Figure~\ref{fig:particles}, we have two particles whose trajectories
intersect.  Before and after crossing the two particle
Lagrangian~\ref{eq:lag} describes the system, but we must give a
prescription for determining the the mass parameter $\tilde{M_1}$ that
determines the geometry between the particles after crossing.
($M_0/\lambda$ and $M_2/\lambda$ are unchanged because the former is
the mass of the background black hole and latter is the total
conserved energy.)  The splicing prescription is derived from the
observation that the crossing of particles does not involve any actual
displacement so that we expect total energy and momentum to be
conserved.  Applying this to the right hand Figure~\ref{fig:particles}
and using the effective momenta given in Equation~\ref{eq:can} we
arrive at the splicing condition:
\begin{equation}
\left(e^{2\lambda r_c} - \frac{M_0\pi}{\lambda^2}\right)
\left(e^{2\lambda r_c} - \frac{M_2\pi}{\lambda^2}\right)
=
\left(e^{2\lambda r_c} - \frac{M_1\pi}{\lambda^2}\right)
\left(e^{2\lambda r_c} - \frac{\tilde{M}_1\pi}{\lambda^2}\right)
\end{equation}
where $r_c$ is the position at which the crossing occurs.  Letting
$M_b/\lambda = (M_2 - M_1)/\lambda$ and $\tilde{M}_b/\lambda =
(\tilde{M}_1 - M_0)/\lambda$ be the energies of the infalling particle
before and after collision with the outgoer we can write the splicing
condition as:
\begin{equation}
\frac{\tilde{M}_b\pi}{\lambda^2} =
\frac{M_b\pi}{\lambda^2}  + 
\frac{ (M_b\pi/\lambda^2) (M_a\pi/\lambda^2) }
     {\left(e^{\lambda r_c} + \sqrt{M_1\pi}/\lambda\right)
      \left(e^{\lambda r_c} - \sqrt{M_1\pi}/\lambda\right)}
\label{eq:bsplice}
\end{equation}
Here $M_a/\lambda$ is the energy of the outgoing particle prior to the
collision.   We see that the energy of the infaller is shifted by the
collision.   The significance of this shift will be discussed in
Section~\ref{sec:break}.

\subsection{Equations of Motion}
\label{sec:eom}
To complete the description of the dressed classical mechanics of
particles in dilaton gravity we must compute the equations of motion
since the classical trajectories will be necessary for the WKB
quantization of the theory in the next section. Since we are in the
Hamiltonian formalism the equations of motion are given by $\dot{r}_i
= \partial H/\partial p_i = (1/\lambda) \partial M_n/\partial p_i$
where all the $r_i$ and the $p_j$ for $j \neq i$ are held constant in
taking the partial derivatives.  Generically, the classical kinematics
of a highly nonlinear Lagrangian like the one in Equation~\ref{eq:lag}
can only be integrated if there are a sufficient number of conserved
quantities present in the system.  Fortunately, because neither the
dilaton nor the graviton is dynamical we expect that the energy of
every particle ($E_i = (M_i - M_{i-1})/\lambda$) is individually
conserved in the absence of crossing of trajectories.  

It is shown in Appendix~\ref{sec:integ} that the system of $n$ dressed
particles is classically integrable since the $M_i$ are
time-independent and have mutually vanishing Poisson brackets.
%(In fact, the $M_i$ are mutually
%commuting even if the momenta and coordinates are promoted to
%operators so that the system is quantum mechanically integrable in the
%sense that the n-particle Fock space has N conserved, mutually
%commuting operators.)  
With this observation in hand, we derive the equations of motion by
noting from Equation~\ref{eq:can} that the Hamiltonian $M_n/\lambda$
on depends on $p_i$ only via its dependence on $M_{n-1}$ which in turn
depends on $M_{n-2}$ and so on until we reach $M_i$ which is
implicitly expressed as a function of $p_i$.  So we have $\dot{r}_i =
(\partial M_n/\partial M_{n-1})_{r_i,p_i}\cdots(\partial
M_{i+1}/\partial M_i)_{r_{i+1},p_{i+1}} (\partial M_i/\partial
p_i)_{r_i,M_{i-1}}$.  These derivatives can be computed by
differentiating both sides of Equation~\ref{eq:can} and rearranging
terms. The subscripts indicate the variables in Equation~\ref{eq:can}
that are to be held constant while taking the derivatives.  Putting
everything together we find the equations of motion for the dressed
particles:
\begin{equation}
\label{eq:eom}
\dot{r}_i = \left(\eta_i - \frac{\sqrt{M_i \pi}}{\lambda e^{\lambda
r_i}} \right) 
\frac{e^{\lambda r_n} - \eta_n \sqrt{M_n\pi}/\lambda}{
      e^{\lambda r_{i+1}} - \eta_{i+1} \sqrt{M_i\pi}/\lambda}
\prod_{k=i+1}^{n-1} \left( 
\frac{e^{\lambda r_k} - \eta_k \sqrt{M_k \pi}/\lambda}{
      e^{\lambda r_{k+1}} - \eta_{k+1} \sqrt{M_k \pi}/\lambda}
\right)
\end{equation}
The integrability of the system now comes to our rescue - all the
$M_i$ are constant and so we can integrate the equation of motion of
the outermost particle, use the trajectory to integrate the motion
of the next inner particle and so on until all the classical
trajectories have been computed.   If particle trajectories cross we
must use one set of $M_i$ prior to crossing and another after crossing
following the splicing prescription described in the previous subsection.
For use in the next section we integrate the equation of motion of a
single dressed particle:
\begin{equation}
\label{eq:1motion}
\dot{r} = \eta - \frac{\sqrt{M_1 \pi}}{\lambda e^{\lambda r}}  \: \:
\: \:  ;
\: \: \: \:
(e^{\lambda r} - \eta \sqrt{M_1 \pi}/\lambda ) =
e^{\eta\lambda t}  (e^{\lambda r(0)} - \eta \sqrt{M_1 \pi}/\lambda ) 
\end{equation}
This equation of motion is identical to the equation for geodesic
motion of a massless particle in the metric~(\ref{eq:met1}) for a
black hole of mass $M_1/\lambda$.  Indeed, for an outgoing trajectory
($\eta = +1$) we see that the particle can only escape to infinity if
$e^{\lambda r(0)} > \sqrt{M_1 \pi}/\lambda$.  As discussed in
Section~\ref{sec:classic} a black hole of mass $M_1/\lambda$ has its
horizon at $e^{\lambda r} = \sqrt{M_1 \pi}/\lambda$.  We see that at
the level of the equations of motion a single dressed particle moves
as though the metric is determined by the total ADM mass ($M_1$) as
opposed to just the mass of the background black hole ($M_0$).

We would like to use the effective Lagrangian in Equation~\ref{eq:lag}
to construct the quantum mechanics of the system.  Given $N$ particles
we have a dressed Lagrangian that describes the propagation of these
particles within $N$ particle Fock space.  We need not be disturbed
that the lack of a second quantized formulation will prevent us from
seeing the particle production associated with the Hawking radiation.
The Hawking flux is not associated with ``vertices'' in the usual
perturbative sense of particle production - the Hawking particles
appear because of a mismatch of vacua between the inertial pbserver at
the horizon and the observer at inifinty.  Given that we are in the
N-particle Fock space we do not expect the number of particles to
change via interactions between the $N$ particles.  Therefore, given
that $N$ Hawking particles are produced, we can study their back
reaction effects by constructing the quantum mechanics described by
the Lagrangian~\ref{eq:lag} within the $N$ particle Fock space.
Nevertheless, there is a difficulty with quantizing the system of
gravitationally dressed particles.  Because the Lagrangian is so
nonlinear in the coordinates and momenta we cannot simply promote
these quantities to operators and canonically quantize the system
since we would be faced with difficult normal ordering problems.  The
most reasonable procedure towards quantizing the system appears to be
to study the semiclassical limit in which $\hbar \rightarrow 0$ so
that action of classical trajectories of the particles and the small
fluctuations around these trajectories dominate the quantum physics.
In the next section we will describe the WKB procedure for carrying
out such a quantization and then we will apply the procedure to
computing the wavefunctions of the dressed particles.  The resulting
states presumably include all the self-interaction corrections arising
from self-exchange of the longitudinal graviton and dilaton.

\section{Computation of Dressed Wavefunctions}
\label{sec:dressing}
We are forced to resort to semiclassical methods to quantize the
effective theory because it is too nonlinear for canonical or full
path integral quantization.  The basic idea of the semiclassical
method is to observe that in the limit of small $\hbar$, the transport
of wavefunctions is dominated by classical trajectories. (See
Gutzwiller~\cite{gutz}.)  In general, if $\psi(r(0)) = \rho(r(0))
\exp{iS(r(0))}$ is the wavefunction at early times, the transport of
the wavefunction to late times is described by the equation $\psi(r,t)
= \int dr(0) \psi(r(0)) K(r(0);r,t)$.  Here $K$ is the propagator
given by the path integral $K(r(0);r,t) = \int_{r(0)}^{r(t)} {\cal
D}{\bf r} \exp{(i/\hbar) S({\bf r})}$ and $S$ is the action for the
path ${\bf r}$.  For small $\hbar$, extrema of the action dominate the
path integral and the integral can be well approximated as the
saddlepoint phase times the determinant of the quadratic fluctations
around the saddlepoint.  To use the semiclassical propagator to
calculate the transport of wavefunctions we must calculate the above
integral over the initial positions $r(0)$.  Because the propagator
has been calculated by the method of saddlepoints, it is only
consistent to do the integral over $r(0)$ by the method of
saddlepoints as well.  It is useful to note that the phase of the
integrand is proportional to $1/\hbar$ whereas the amplitude is $O(1)$
so that the saddlepoint is consistently calculated from the phases
alone.  In other words, we evaluate the integrand at the coordinate
$r(0)\myprime$ at which the phase of the intgrand is minimized and
then we calculate the determinant of quadratic fluctuations around
this point.  Because it is suppressed by a power of $\hbar$, the
amplitude $\rho$ does not enter the calculation of the determinant and
is simply evaluated at the saddlepoint.  Putting all these facts
together we arrive at the following expression for the semiclassical
propagation of a wavefunction that starts at $t = 0$ as $\psi_0(r(0))
= \rho(r(0)) e^{\frac{i}{\hbar} S_0(r(0))}$:
\begin{eqnarray}
\psi(r,t) &=& \rho(r_0) \sqrt{ \frac{|-\partial^2 R(r,r_0,t)/\partial r
\partial r_0|}
{-\left( 
\partial^2 S_0(r_0)/\partial r_0^2 +
\partial^2 R(r,r_0,t)/\partial r_0^2 \right)}}
e^{\frac{i}{\hbar} (S_0(r_0) + R(r,r_0,t))}
 \nonumber \\
R(r,r_0,t) &=& \int_{r_0}^r p(r\myprime,E) dr\myprime - Et
\label{eq:wkb}
\end{eqnarray}
In this expression $E$ and $r_0$ are the self consistent solutions to
the equations:
\begin{equation}
\label{eq:self}
r = r(r_0,E,t) \: \: \: ; \: \: \: \frac{\partial S_0}{\partial r_0} =
p(r_0, E)
\end{equation}
The first of these equations is the equation of motion for the
classical trajectory of energy $E$ that starts at $r_0$ and the second
equation is standard relation between the action and the momentum of a
classical particle.  The semiclassical method requires that the
frequency of the wavefunction varies sufficiently slowly for us to
define a local energy in every segment of the wave.  The content of
the above equations is that this local energy at $r_0$ is transported
along the classical trajectory of that energy that starts at $r_0$ and
is deposited at the final point of the trajectory after a time $t$.
%During the transport, some energy may be lost or gained in the process
%of moving through a background potential.  
As a final check note that
imposing the condition that $E$ is constant giving an energy
eigenstate can be shown to yield the following wavefunction:
\begin{equation}
\label{eq:eeig}
\psi(r,t) = \frac{1}{\sqrt{\dot{r}}} e^{\frac{i}{\hbar}\left[
\int^r p(r,E) \,dr - Et \right]}
\end{equation}
This is the familiar WKB expression for energy eigenstates in one
dimensional quantum mechanics.  We now apply this formalism to the
gravitational self-interaction of particles in dilaton gravity.  Note
that in computing overlap integrals of semiclassical wavefunctions it
is only consistent to use the method of saddlepoints since the WKB
wavefunctions have themselves been computed in that manner.

\subsection{Self-Interaction Corrections To Radiation}
\label{sec:selfcorr}
We will now compute the self-interaction corrections to wavefunctions
of single particles propagating in dilaton gravity. Having computed
the dressed spectrum we will repeat the calculation of the Hawking
radiation in Section~\ref{sec:classic} with the improved modes.  In
fact we will only work with the phase of the wavefunction because, as
discussed above, the saddlepoints in the overlap integrals that give
the Bogliubov coefficients are insensitive to the amplitude which is
suppressed by a power of $\hbar$.  Equation~\ref{eq:wkb} says that the
phase of the single particle dressed WKB wavefunction is given by
$S(r,t) = S_0(r_0) + \int_{r_0}^r p_c dr - (M_1 - M_0)t/\lambda$ where
we have identified the energy of the particle as the ADM mass
($M_1/\lambda$) minus the mass of the background black hole, and where
$p_c$ is the coefficient of $\dot{r}$ in Equation~\ref{eq:1part}.  We
are interested in the self-interaction of the outgoing wavefunctions
that are responsible for producing the Hawking radiation. Integrating
$p_c$ we find:
\begin{eqnarray}
R = \int p_c \, dr = &  \frac{1}{\pi} \left[  e^{\lambda r} \left(
\frac{\sqrt{M_0\pi}}{\lambda} - \frac{\sqrt{M_1\pi}}{\lambda} \right) +
 \left( \frac{M_1\pi}{\lambda^2} - e^{2\lambda r} \right)
  \ln\left(e^{\lambda r} - \frac{\sqrt{M_1\pi}}{\lambda} \right)   \right.
-  \nonumber \\
& \left. \left( \frac{M_0\pi}{\lambda^2} - e^{2\lambda r} \right)
  \ln\left(e^{\lambda r} - \frac{\sqrt{M_0\pi}}{\lambda} \right) \right]
\label{eq:R}
\end{eqnarray}
According to Equation~\ref{eq:self} we must also self consistently
solve for $M_1$ and $r_0$ using the equation of motion
(Equation~\ref{eq:1motion} ) and the initial condition $S_0$.  We will
imagine that the self-interactions are turned off until $t=0$ and the
wavefunction starts in an outgoing Kruskal eigenstate as in
Equation~\ref{eq:kruskalout}.  So we take $S_0 = (2k/\pi) (e^{\lambda
r} - \sqrt{M_0\pi}/\lambda)$.    The self-consistency conditions in
Equation~\ref{eq:self} are solved in Appendix~\ref{sec:onewave} 
to arrive at the  leading order dressed WKB phase at late times:
\begin{equation}
\label{eq:leading} 
S(r,t) = \frac{2}{\pi} \frac{\sqrt{M_0 \pi}}{\lambda}
\left(\exp\left(\frac{k\lambda}{\sqrt{M_0 \pi}}\right) - 1 \right)
e^{-\lambda t}
\left(e^{\lambda r} - \frac{\sqrt{M_0\pi}}{\lambda} \right)
\end{equation}
The late time wavefunction is $e^{iS}$ and therefore looks exactly
like the Kruskal eigenstate in the initial condition with a redefined
frequency.  Note that $(\sqrt{M_0\pi}/\lambda)
(\exp(k\lambda/\sqrt{M_0\pi}) -1)$ is $k$ to leading order for small
$k$ so that states of small Kruskal momentum are unaffected.  However,
as we shall shortly see, virtual states of very large Kruskal momentum
are responsible for the production of the hawking radiation and so
this frequency redefinition is important for them.
    
Having computed the one-particle dressed wavefunctions we are in a
position to analyze the physical effects arising from gravitational
self-interaction.  Kraus and Wilczek have found that such
self-interactions cause a energy dependent shift of the radiation
temperature for Schwarschild and Reissner-Nordstrom black holes
(\cite{kw1}, \cite{kw2}) and it is interesting to understand whether
such effects can occur in the context of the two dimensional black
holes.   We have computed the dressed analogues of the Kruskal
eigenstates above, and if we also compute the dressed energy
eigenstates we can find the Bogliubov transformations responsible for
producing the Hawking radiation.   From Equations~\ref{eq:eeig}
and~\ref{eq:R} we know that the phase of a dressed energy eigenstate
is $S_\omega(r,t) = R(r) - \omega t$ with $(M_1 - M_0) / \lambda \equiv
\omega$.   In computing the dressed Kruskal eigenstates we assumed
that the particles had small energies and expanded to leading order in
these energies.   Under the same assumptions, $S_\omega$ becomes:
\begin{equation}
S_\omega(r,t) = \frac{\omega}{\lambda}\ln\left(e^{\lambda r} -
\frac{\sqrt{M_0\pi}}{\lambda} \right) -\omega t - \frac{\omega}{2\lambda} 
\end{equation}
Writing the dressed energy eigenstate as $\phi_\omega = e^{iS_\omega}$
(we are ignoring amplitudes) we can compute the Bogliubov
tranformations between the dressed states using the
relations~(\ref{eq:bogints}).  To be consistent within the semiclassical
framework, the integrals in Equation~\ref{eq:bogints} should be
regarded as overlap integrals between the two dressed bases and should
be carried out by the method of saddlepoints.  Define
$\omega\myprime/\lambda = (2/\pi)
(\sqrt{M_0\pi}/\lambda)[\exp(k\lambda/\sqrt{M_0\pi}) - 1]$ 
and $u_f = e^{-\lambda t} (e^{\lambda r} - \sqrt{M_0 \pi}/\lambda)$ so
that the dressed Kruskal states are $\psi =
\exp(i(\omega\myprime/\lambda) u_f)$. Then the saddlepoints for the
$\alpha$ and $\beta$ integrals occur at:
\begin{equation}
\alpha^*: \: e^{-\lambda t} = \frac{\omega}{\omega\myprime(e^{\lambda
r} - \frac{\sqrt{M_0\pi}}{\lambda})} \: \: \: ; \: \: \:
\beta^*: \: e^{-\lambda t} = - \frac{\omega}{\omega\myprime(e^{\lambda
r} - \frac{\sqrt{M_0\pi}}{\lambda})}
\end{equation}
We are interested in the radiation of moderate $\omega$ seen by the
asymptotic observer and so we see that the saddlepoint for $\alpha$
occurs at very late times.  The saddlepoint time for $\beta$ is off
the real axis, but also has a large real part.  Because of this it is 
consistent to use the late time form of the dressed wavefunctions as
we are doing. Ignoring amplitudes of wavefunctions, the saddlepoint
integral gives:
\begin{eqnarray}
\alpha^*_{\omega\omega\myprime} = e^{i (3\omega/2\lambda)}
\left(\frac{\omega\myprime}{\omega}\right)^{\frac{i\omega}{\lambda}}
\: \: \: ; \: \: \:
\beta^*_{\omega\omega\myprime} = e^{i (3\omega/2\lambda)}
\left(\frac{-\omega\myprime}{\omega}\right)^{\frac{i\omega}{\lambda}}
= \alpha^*_{\omega\omega\myprime} e^{\frac{-\pi\omega}{\lambda}}
\end{eqnarray}
As in Section~\ref{sec:classic} $|\alpha_{\omega k}/\beta_{\omega k}|$
is independent of $k$ and so we reach the conclusion once again that
the dressed energy eigenstates are thermally populated with a
temperature of $(2\pi)/\lambda$ if the system is prepared in a state
annihilated by the dressed Kruskal states.  

We have not found a shift of the Hawking temperature unlike Kraus and
Wilczek (\cite{kw2,kw3}).  This is not surprising in retrospect
because the temperature of dilaton black holes depends only on the
cosmological constant and we do not expect the self-interactions of
the matter particles to renormalize this quantity.  Nonetheless, the
frequency redefinition derived here is quite dramatic and we may
wonder whether there are any physical effects associated with it.  To
resolve this question we would need to compute the amplitudes of the
effective wavefunctions and examine whether the n-point functions of
the Hawking state are changed.  We do not expect any such changes
because the frequency renormalization derived here does not mix
positive and negative frequencies.  The structure of the effective
wavefunctions strongly suggests that the net effect of the
self-interactions is merely a frequency redefinition.

In computing the Bogliubov tranformations we used only the leading
order dressed wavefunctions in Equation~\ref{eq:leading}.  One may
wonder whether the subleading corrections could lead to nonthermal
population of states in the radiation.  We do not expect this to
happen because the computations in Appendix~\ref{sec:onewave} show
that the Kruskal eigenstates will depend on time via factors of
$e^{-\lambda t}$.  Consequently these states are periodic in imaginary
time with a period of $2\pi/\lambda$ suggesting that the spectrum will
look thermally populated even when the subleading terms are accounted
for.  

Before concluding this section there are two more useful lessons to
draw.  The first is that the Kruskal momenta contributing to late time
radiation grow linearly with time and the second is that the
self-interaction corrections are unimportant for infalling states.  To
show the first point we return to Equation~\ref{eq:bog1} to express the
dressed energy eigenstates in terms of the dressed Kruskal states:
$\phi_{\omega} = \int d\omega\myprime (\alpha_{\omega\omega\myprime}
\psi_{\omega\myprime} + \beta_{\omega\omega\myprime}
\psi_{\omega\myprime}^* )$.  The $\omega\prime$ dependent part in
integrand of the first term is
$\exp(-i(\omega/\lambda){\ln{\omega\myprime}} +
i(\omega\myprime/\lambda) u_f)$ ($u_f$ is defined above).  The
saddlepoint that dominates the integral over $\omega\prime$ is
at $\omega\myprime = \omega/u_f = \omega/(e^{-\lambda
t}(e^{\lambda r} - \sqrt{M_0 \pi}/\lambda))$ which grows exponentially
in time at fixed $r$.   Since $\omega\myprime \propto
\exp(k\lambda/\sqrt{M_0\pi})$ for large $k$, we see that $k$ grows
linearly in time.\footnote{There is a subtlety regarding
whether the $\phi_\omega$ is to be expressed as an integral over $k$
or over the redefined frequency $\omega\myprime =
\sqrt{M_0\pi}[\exp(k\lambda/\sqrt{M_0\pi}) - 1]$.   In the above
discussion we used the redefined frequency in the integration measure.
In fact in the small $\hbar$ limit of interest to us here (we are
considering the limit where the semiclassical wavefunctions are reliable) 
both integration measures yield the same conclusion concerning the
values of $k$ that dominate the integral.  This is because, in the
semiclassical limit, the integral is dominated by the $O(1/\hbar)$
phases in the wavefunctions $\psi$ and in the Bogliubov coefficients.
This means that the $O(1)$ measure in the integral is not important for
determining the important regions of the integrand.}

Finally, we can ask whether incoming states are strongly affected by
self-interactions.  Physically, we do do not expect large effects
because the relationship between momentum and energy for incoming
particles does not become singular at horizon.  The lack of any large
effects can be shown in the WKB lanuguage by taking $S_0 =
(2k/\pi)(e^{\lambda r} + \sqrt{M_0\pi}/\lambda)$, so that we have an
incoming Kruskal eigenstate as the initial condition, with $\eta = -1$
in Equation~\ref{eq:can} for the canonical momentum.  In the
calculation of the dressed wavefunctions in Appendix~\ref{sec:onewave},
the exponentiation of of the Kruskal frequency $k$ arises in the
course of solving the WKB consistency conditions in
Equation~\ref{eq:self}.  It is a simple matter to show that the
incoming initial conditions described above do not result in such an
exponentiation for low energy incoming particles even in the region
near the black hole horizon.

%The basic reason that the outgoing state was strongly affected by the
%self-interactions was Equation~\ref{eq:iter2} relating $\epsilon$ to
%the exponential of $k$.  This followed from the relation $\partial
%S_0(r_0)/\partial r_0 = p_c(r_0)$ with the initial $S_0$ chosen to be
%an outgoing Kruskal state and $p_c(r_0)$ being the dressed canonical
%momentum of an outgoing particle.  We now take $S_0 =
%(2k/\pi)(e^{\lambda r} + \sqrt{M_0\pi}/\lambda)$ so that we have an
%{\it incoming} Kruskal eigenstate to start with and we take $\eta =
%-1$ in Equation~\ref{eq:can}.  Then defining $\epsilon$ and $\delta$
%as before we find that:
%\begin{equation}
%k + \epsilon = \lambda \left(\frac{\sqrt{M_0 \pi}}{\lambda} +
%\epsilon + \delta \right) \ln\left[
%1 + \frac{\epsilon}{2\sqrt{M_0 \pi}/\lambda + \epsilon + \delta} 
%\right]
%\end{equation}
%It is clear from this equation that even close to the horizon ($\delta
%\ll \sqrt{M_0\pi}/\lambda$), for small energies ($\epsilon \ll
%\sqrt{M_0\pi}/\lambda$) $\epsilon$ is linearly related to $k$ in the
%leading order.  So we will not find the frequency redefinition that
%appeared in the outgoing case.  

\section{In-Out Scattering}
\label{sec:io}
%Having understood the gravitational self-interactions of particles in
%dilaton gravity we can move on to the even more interesting situation
%where outgoing flux emerges from the dynamical background produced by
%an infalling quantum state.  
Giddings and Nelson have carried out a calculation of the Hawking flux
in the $1+1$ dilaton geometry produced by a classical infalling state
(\cite{gn}).  In this section we will study the radiation produced in
the dynamical background of a quantum mechanical infalling state.  A
complete treatment of this problem involves consideration of the
two-particle states in Fock space in which one particle is coming in
and one particle is going out, treated with the two particle version
of Lagrangian~(\ref{eq:lag}).  The full problem is very hard to
analyze because of the non-linearities of the gravitational
interactions between the particles and because the self-interactions
do not commute with the in-out scattering.  We will consider the
structure of the states in the full two particle Fock space in the
next section and restrict ourselves to an approximate (and possibly
more insightful) version of the problem here.  First of all, we will
semiclassically evolve an infalling quantum mechanical packet to
produce a geometry whose ADM mass is uncertain and then we will study
the Hawking problem in this background.
%  To study this we consider a scenario in which
%there is one incoming and one outgoing particle at early times
%described by the two particle version of the Lagrangian~\ref{eq:lag}.
%(See the diagram on the right in Figure~\ref{fig:particles}.)  
By examining the form of the resulting scattering of the outgoing
radiation we are able to identify the in-out scattering matrix for an
arbitrary infalling state up to certain ordering problems.  This will
enable us to identify the effect an arbitrary infalling state has on
the spectrum of outgoing particles.  From this identification we are
able to draw some qualitative conclusions regarding the return of
information in the radiation in the presence of the full
self-interactions.

We now want to construct two-particle WKB wavefunctions.  This can be
done in the same way as discussed at the beginning of
Section~\ref{sec:dressing}.  The phase of the two particle
wavefunction is given by $S(r_a, r_b, t_a, t_b) = S_0(r_a(0),r_b(0)) +
\int p_a dr_a + \int p_b dr_b - E_a t_a - E_b t_b$.  In this equation
we have two times $t_a$ and $t_b$ because we may choose to evolve the
incoming part of the two particle wavefunction forward for a different
amount of time from the outgoing part of the wavefunction.  As in the
one particle case we must solve self-consistently for $E_a, E_b,
r_a(0)$ and $r_b(0)$ in terms of $r_a, r_b, t_a$ and $t_b$ by using
the two equations of motion and the initial conditions that $\partial
S_0(r_a(0),r_b(0))/\partial r_a(0) = p_a(r_a(0))$ and $\partial
S_0(r_a(0),r_b(0))/\partial r_b(0) = p_b(r_b(0))$.  In this section we
are interested in the effect an incoming state has on the outgoing
radiation.  So we will take $t_b = 0$ so that the incoming state is
not evolved forward in time at all - we simply specify the incoming
state at $t=0$ and compute its effects on the outgoing part of the
wavefunction.  In other words, instead of specifying the two particle
configuration space as $\{r_a, r_b\}$ at some time $t$ we choose the
configuration space to be $\{r_a, r_b(0)\}$ and write the wavefunction
in terms of these coordinates.  Furthermore we will prepare the system
at $t=0$ so that $S_0(r_a(0),r_b(0)) = S_{0a}(r_a(0)) +
S_{0b}(r_b(0))$ - in other words, the incoming and outgoing
wavefunctions are not entangled at $t=0$ either by assuming that the
interactions are turned off at earlier times or by picking well
separated initial wavepackets.   As in Section~\ref{sec:dressing}
wavefunction amplitudes in the semiclassical approximation are
computed as determinants of fluctuations around saddlepoints and are
relatively unimportant since they are suppressed by a power of
$\hbar$.  For this reason we will omit them as we did in the previous
section.

\subsection{Leading Order In-Out Scattering}
\label{sec:leading}
We begin by turning off the self-interactions in order to understand
the effects of the in-out interactions by themselves.  The effects of
self-interactions are considered in Section~\ref{sec:ioself}.

Consider an incoming quantum wavepacket propagating in the background
of a black hole of mass $M_0/\lambda$.  The state is specified by an
initial wavefunction $\psi_{b0} = \rho_0 \exp{iS_{b0}(r_b(0))}$.
Removing the self-interactions amounts to keeping only the tadpole
diagram proportional to $E$ in Figure~\ref{fig:diags} and so we will
call this the ``tadpole approximation''. As discussed in
Section~\ref{sec:noninter} this can be achieved by linearizing the
relation between the momentum and energy of a particle in
Equation~\ref{eq:can}.  Doing this gives the following relations:
\begin{equation}
p_b = -\frac{M_b}{\lambda} \frac{e^{\lambda r_b}}{e^{\lambda r_b} +
\frac{M_0\pi}{\lambda}} \: \: ; \: \:
\dot{r}_b = -1 - \frac{\sqrt{M_0\pi}}{e^{\lambda r_b} \lambda} 
\label{eq:bapprox}
\end{equation}
Integrating this equation of motion we find that the classical
trajectories dominating the evolution of the infalling wavepacket have
the equations:
\begin{equation}
\left(e^{\lambda r_b} + \frac{\sqrt{M_0\pi}}{\lambda} \right) =
e^{-\lambda t}
\left(e^{\lambda r_b(0)} + \frac{\sqrt{M_0\pi}}{\lambda} \right)
\label{eq:btraj}
\end{equation}
If the the initial wavepacket is not an energy eigenstate, then the
geometry will not be in an eigenstate of the ADM mass.  Indeed,
identifying $p_{b0} = \partial S_{b0}/\partial r_b(0)$ as the momentum
of the initial state, the relations in Equation~\ref{eq:bapprox} tell
us that the sector of the wavefunction dominated by the classical
trajectory starting at $r_b(0)$ has an ADM mass given by:
\begin{equation}
\frac{M_2}{\lambda} \equiv
\frac{M_0 + M_b}{\lambda} = \frac{M_0}{\lambda} + e^{-\lambda r_b(0)}
\left(e^{\lambda r_b(0)} + \frac{\sqrt{M_0\pi}}{\lambda}\right)
\left(- \frac{\partial S_{b0}}{\partial r_b(0)}\right)
\label{eq:bmass}
\end{equation}
We can use these observations to carry out the semiclassical
evolution of outgoing states in the quantum background produced by the
state $\psi_{b0}$.

Consider an outgoing wavefunction propagating in the background
produced by the infalling packet.   Prior to crossing the infalling
state the outgoer travels in a background of mass $M_0/\lambda$ and
after crossing the background has a mass $M_2/\lambda$ which depends
on the sector of the infalling state under consideration.  Removing
the self-interactions of the outgoer by linearizing
Equation~\ref{eq:can} gives:
\begin{equation}
before:  p_a = \frac{M_a}{\lambda} \frac{e^{\lambda r_a}}{e^{\lambda
r_a} - \frac{\sqrt{M_0\pi}}{\lambda}} \: \: \: ; \: \: \:
after: p_a = \frac{\tilde{M}_a}{\lambda} \frac{e^{\lambda r_a}}{e^{\lambda
r_a} - \frac{\sqrt{M_2\pi}}{\lambda}}
\label{eq:ps}
\end{equation}
where $M_a/\lambda$ is the energy of the trajectory before the
collision and $\tilde{M}_a/\lambda$ is the energy after the collision.
The equations of motion before and after crossing are:
\begin{eqnarray}
before:& \:  
\dot{r}_a = \left(1 -  \frac{\sqrt{M_0\pi}}{e^{\lambda r_a}\lambda}\right) 
\: \:  ; \: \:
e^{\lambda r_a} -\frac{\sqrt{M_0\pi}}{\lambda}  =
e^{\lambda t }\left(e^{\lambda r_a(0)} -\frac{\sqrt{M_0\pi}}{\lambda} \right)
 \label{eq:before} \\
after:& \:  \dot{r}_a = 1 -
\frac{\sqrt{M_2 \pi}}{e^{\lambda r_a}\lambda}
\: \: \: ; \: \: \:
\left(e^{\lambda r_a} -\frac{\sqrt{M_2 \pi}}{\lambda} \right) =
e^{-\lambda(t-t_c)}\left(e^{\lambda r_c}
-\frac{\sqrt{M_2 \pi}}{\lambda} \right) \label{eq:after}
\end{eqnarray}
where $t_c$ and $r_c$ are the time and position at which the outgoing
classical trajectory intersects with the incoming classical trajectory
that starts at $r_b(0)$.  This completes the specification of the
leading order classical mechanics of the outgoing particles in the
backgrounds produced by the infalling quantum state.  These
ingredients will be used in the next section to compute the
approximate semiclassical quantum mechanics of the outgoing states in
the quantum infalling background.

\subsection{Tadpole Corrected Two-Particle Wavefunctions}
\label{sec:tad2part}
As discussed earlier the phase of the two particle wavefunction is
given by:
\begin{eqnarray}
S(r_a,t, r_b(0)) = S_{b0}(r_b(0)) &+& S_{a0}(r_a(0)) + \int_{r_a(0)}^{r_c} p_a
dr_a \nonumber \\
&+& \int^{r_a}_{r_c} p_a dr_a - M_a t_c/\lambda -
\tilde{M}_a(t - t_c)/\lambda
\end{eqnarray}
(We have split the usual integral $\int (p \,dr - H \, dt)$ into the
parts before and after crossing.)  Performing the integrals we find:
\begin{equation}
\int p_a dr_a = \frac{\tilde{M}_a}{\lambda^2} \ln\left[\frac{e^{\lambda r_a} -
\sqrt{M_2\pi}/\lambda}{e^{\lambda r_c} -
\sqrt{M_2\pi}/\lambda}\right] +
\frac{M_a}{\lambda^2} \ln\left[\frac{e^{\lambda r_c} -
\sqrt{M_0 \pi}/\lambda}{e^{\lambda r_a(0)} -
\sqrt{M_0 \pi}/\lambda }\right]
\label{eq:ph1}
\end{equation}
Note that the equations of motion~\ref{eq:before} and ~\ref{eq:after}
tell us the ratios within the logarithms in terms of the time
travelled before and after the collision with the infaller.  Using
these relations in Equation~\ref{eq:ph1} we find that the phase of the
two particle wavefunction in the tadpole approximation is given by:
\begin{equation}
S(r_a,t,r_b(0)) = S_{b0}(r_b(0)) + S_{a0}(r_a(0))
\label{eq:simple}
\end{equation}
So we need only solve for $r_a(0)$ as a function of $r_a$, $t$ and
$r_b(0)$ in order to compute the WKB phase of the two-particle
wavefunction.

      Let us take $S_{a0} = (k/\lambda) (e^{\lambda r_a(0)} -
\sqrt{M_0\pi}/\lambda)$ so that we have an outgoing Kruskal
eigenstate.  Using the equations of motion of the outgoing particle
before and after scattering it is easy to write $r_a(0)$ in terms of
$r_a$, $t$ and $r_c$ where $r_c$ is the position at which the
collision with the infaller occurs.  Next we use the equations of
motion of the infaller before scattering along with the equation of
motion of the outgoer after scattering to find that:
\begin{equation}
\left(e^{\lambda r_c} + \frac{\sqrt{M_0 \pi}}{\lambda} \right)
\left(e^{\lambda r_c} - \frac{\sqrt{M_2 \pi}}{\lambda} \right)
= u_{out} v_{in}
\end{equation}
where we have defined $u_{out} = e^{-\lambda t} (e^{\lambda r_a}  -
\sqrt{M_2\pi}/\lambda)$ and $v_{in} = (e^{\lambda r_b(0)}  +
\sqrt{M_0\pi}/\lambda)$.   At late times on ${\cal I}^+$, which is the
region of interest to us, the solution to the equation is:
\begin{equation}
e^{\lambda r_c} \approx \frac{\sqrt{M_2\pi}}{\lambda} + 
\frac{u_{out} v_{in}}{ \sqrt{M_2 \pi}/\lambda  +
\sqrt{M_0 \pi}/\lambda }
\end{equation}
Using this late time solution for $\exp{\lambda r_c}$ in the outgoing
equations of motion yields a solution for $\exp{\lambda r_a(0)}$ that
gives the following result for the phase of the outgoing part of the
wavefunction:
\begin{equation}
S(r_a(0)) = 
\frac{k}{\lambda} e^{-\lambda t}\left(e^{\lambda r_a} -
\frac{\sqrt{M_2 \pi}}{\lambda} \right)   
+
\frac{\pi}{\lambda^2} \left(k \frac{(M_2 - M_0)/\lambda}{v_{in}}\right)
\end{equation}
Finally,  we can use Equation~\ref{eq:bmass} for the mass of the
geometry $M_2/\lambda$ in the sector of the two-particle wavefunction
labelled by $r_b(0)$.   The semiclassical method is only valid when
the energy of the incoming state is small compared to the mass of the
black hole ($M_b \ll M_0$) so that we can expand the $\sqrt{M_2\pi}$
to leading order arriving at the following expression for the phase of
the two-particle WKB wavefunction:
\begin{eqnarray}
S(r_a,t,r_b(0)) = & S_{b0}(r_b(0)) + \frac{P_-}{\lambda}
e^{-\lambda t}\left(e^{\lambda r_a} -
\frac{\sqrt{M_0 \pi}}{\lambda} \right)   + \nonumber \\
&
\frac{\pi}{\lambda^2} P_-
\left[
1 - \frac{\lambda e^{-\lambda t} v_{in}}{2\sqrt{M_0\pi}} 
\right] P_+ \label{eq:2ph}
\end{eqnarray}
where we have identified $P_- \equiv k$ as the outgoing Kruskal
momentum and $P_+ = M_b/v_{in} = \exp{-\lambda r_b(0)} (-\partial
S_{b0}/\partial r_b(0))$ as the Kruskal momentum flowing in from the
initial coordinate $r_b(0)$.  Comparing with
Equation~\ref{eq:kruskalout} for the outgoing Kruskal eigenstates we
see that the last expression has the structure of a scattering phase
shift induced in the outgoing state by the incoming state.

\subsection{Tadpole Corrected Physics}
First let us examine the structure of the two particle wavefunction in
Equation~\ref{eq:2ph}.   The leading term on the second line,
$\pi P_- P_+/\lambda^2$ is the two dimensional reduction of the
shockwave interaction first advocated by t'Hooft
(\cite{thooft1},\cite{sv21,sv22}).     The term that is suppressed by
$\sqrt{M_0\pi}$ arises because the position of the horizon moves when
the infalling state falls into the black hole.   This shift of the
horizon ``squeezes'' the region outside the black hole causing a
deformation of the wavefunction of states defined in the exterior
region.  In general, the term on the second line represents a
scattering phase shift produced in the outgoing state by the incoming
state and has a structure somewhat reminiscent of the exact S-matrix
in the electromagnetic analogue problem studied by~\cite{peet}.
However, we do not expect the subleading correction computed here to
restore the information about the infalling state to infinity since we
do not have a systematic series of such corrections of the necessary
structure. 

The two particle wavefunction in Equation~\ref{eq:2ph} can be used to
identify the scattering operator that acts on the outgoing state in
the presence of the infalling wavepacket.  To do this observe that in
the WKB method operators are essentially replaced by expectation
values taken locally in every segment of the wavefunction with slowly
varying phase.  For example, the quantity $P_+ = \exp{-\lambda r_b(0)}
(-\partial S_{b0}/\partial r_b(0))$ is to be understood as an
expectation value of $P_+$ taken locally in the neighbourhood of
$r_b(0)$.  Indeed, identifying the momentum $ p_b(0) = \partial
S_{b0}/\partial r_b(0)$ (and similarly $p_a$) we can promote the
coordinates and momenta appearing in the scattering phase shift
computed above to operators and arrive at the scattering operator
acting on the outgoing states up to ordering problems.  This
scattering operator has the effect of entangling the outgoing and
incoming states.

Next, we calculate the Hawking radiation in the dynamical background
produced by the incoming state via computation of the Bogliubov
coefficients.  In order to do this we need the the form of the energy
eigenstates as measured by the asymptotic observer.  Using
Equation~\ref{eq:eeig} we find the phase of an asymptotic energy
eigenstate to be:
\begin{eqnarray}
S_{\omega}(r_a,t,r_b(0)) &=& S_{b0}(r_b(0)) + \int^r p_a dr_a - \omega t
\nonumber\\ 
&=& 
S_{b0}(r_b(0)) + \frac{\omega}{\lambda} \ln\left(e^{\lambda r_a}
-\frac{\sqrt{M_2\pi}}{\lambda}  \right) - \omega t \label{eq:engeig}
\end{eqnarray}
Note that $S_\omega$ diverges at $\exp{\lambda r_a} =
\sqrt{M_2\pi}/\lambda$ where the horizon of the black hole is located
after the infalling mode has fallen in.  In previous sections we have
only worked with states defined in the exterior of the black
hole but in what follows we will need to augment the wavefunctions
$\exp{i S_\omega}$ with modes defined in the interior of the black
hole.  This is necessary because the Kruskal eigenstates in our
initial conditions have support both inside and outside the black hole
and therefore a full analysis of the final state requires modes
defined in both these regions.  Since the definition of a ``particle''
is somewhat ambiguous inside the black hole, we are free to pick any
convenient basis to augment the set $\phi_\omega = \exp{i
S_\omega}$. In the interior of the black hole we therefore pick the
set $\hat{\phi}_\omega = \exp{i \hat{S}_\omega}$ where:
\begin{equation}
\hat{S}_\omega(r_a,t,r_b(0)) = S_{b0}(r_b(0)) +
\frac{\omega}{\lambda} 
\ln\left(\frac{\sqrt{M_2\pi}}{\lambda} - e^{\lambda r_a}\right)
- \omega t
\end{equation}

In Equation~\ref{eq:bog1} we had defined the Bogliubov coefficients
$\alpha$ and $\beta$ relating $\phi_\omega$ to the Kruskal basis.  Now
define $\hat{\alpha}$ and $\hat{\beta}$ analogously to relate
$\tilde{\phi}_\omega$ to the Kruskal states.  Using the
relations~\ref{eq:bogints} and their analogues for $\hat{\alpha}$ and
$\hat{\beta}$ we find that saddlepoint calculations give the following
results:\footnote{Once again, since we are working in the
semiclassical limit of small $\hbar$ and since the computation of
Bogliubov coefficients is to be regarded as an overlap integral
between semiclassical wavefunctions, it is only consistent to perform
these integrals via the saddlepoint method.}
\begin{eqnarray}
\alpha^*_{\omega k}(r_b(0)) &=&  e^{i\pi k P_+/\lambda^2} 
e^{i\omega/\lambda} \left(\frac{k}{\omega}\right)^{i\omega/\lambda}
 \nonumber \\
\beta^*_{\omega k}(r_b(0)) &=& 
- e^{- i\pi k P_+/\lambda^2} 
e^{i\omega/\lambda} \left(\frac{-k}{\omega}\right)^{i\omega/\lambda}
= - \alpha^*_{\omega k} \, e^{-\pi\omega/\lambda}
\nonumber \\
\hat{\alpha}^*_{\omega k}(r_b(0)) &=& 
\alpha^*_{\omega k}(r_b(0))
\; e^{\pi\omega/\lambda}
\: \: \: \: \: ; \: \: \: \: \:
\hat{\beta}^*_{\omega k}(r_b(0)) = 
\beta^*_{\omega k}(r_b(0)) 
\; e^{\pi\omega/\lambda}
\label{eq:boginout}
\end{eqnarray} 
We have written these Bogliubov coefficients as functions of $r_b(0)$
in order to emphasize that $P_+ = \exp{-\lambda r_b(0)} (-\partial
S_{b0}/\partial r_b(0))$  is a function of $r_b(0)$ and that the
in-out scattering represents an entanglement that prevents us from
separating the Hilbert space into a product of incoming and outgoing
states. 

Despite the entanglement produced by the scattering, since
$|\alpha_{\omega k}/\beta_{\omega k}|$ is independent of both $k$ and
$r_b(0)$, the standard arguments used in Section~\ref{sec:classic} tell us
that the Kruskal vacuum state contains a thermal spectrum of particles
of temperature $2\pi/\lambda$. (See~\cite{gn} and~\cite{bd} for
explanations of the standard arguments.)   Indeed, we can show that
the final Hawking state is left completely unaffected by the $P_+ P_-$
phase accumulated during the in-out scattering.   To show this, let
$a_k$ be the annihilation operator for the Kruskal mode $k$ and let
$b_\omega$ and $\hat{b}_\omega$ be the annihilation operators for the
asymptotic modes $\phi_\omega$ and $\hat{\phi}_\omega$ defined in the
previous section.   Then the Bogliubov coefficients can be used to
relate the various creation and annihilation operators (\cite{gn}):
\begin{equation}
a_k = \int dw \, \left[ 
\alpha_{\omega k} \, b_\omega + \beta^*_{\omega k} \, b^+_\omega
+
\hat{\alpha}_{\omega k} \, \hat{b}_\omega + \hat{\beta}^*_{\omega k}
\, \hat{b}^+_\omega \right]
\label{eq:adef}
\end{equation}
The state of the system is defined to be the one that is annihilated
by all the operators $a_k$. Putting in the Bogliubov coefficients in
Equation~\ref{eq:boginout} it is clear that the scattering phase can
be factored out of Equation~\ref{eq:adef} giving:
\begin{equation}
a_k = e^{-i\pi k P_+/\lambda^2} a_k\myprime
\end{equation}
where $a_k\myprime$ is the Kruskal annihilation operator in the
absence of the scattering phase shifts.   Since the Hawking state in
the absence of scattering is the one annihilated by all the
$a_k\myprime$ we can see that the same state is annihilated by the
$a_k$ so that the in-out scattering does not affect the state at all.
Indeed, the only trace of the infalling state in the radiation is that
the energy eigenstates are defined in a geometry whose mass is
determined by the energy carried into the black hole by the
infaller.

\subsection{In-Out Scattering With Self-Interaction}
\label{sec:ioself}
In previous sections we have computed the effects on the Hawking
radiation of self-interactions and of an infalling state.   The former
led to a renormalization of the frequency of outgoing Kruskal states
and the latter produced a scattering phase shift.   We may wonder
whether the two effects together can be responsible for any added
recovery of information.   The problem is that the severe
non-linearities of the self-interacting system in the presence of an
incoming state preclude analysis except in very special kinematical
regimes.   Therefore, despite extensive calculations in the more
complete scenario we will be satisfied with a qualitative discussion
of the phenomena that can be expected.
    
 The in-out scattering does not commute with the self-interactions
since the scattering shifts the phase of the wavefunction by an amount
proportional to the outgoing Kruskal momentum and the
self-interactions renormalize this momentum.  Let us imagine that the
self-interactions are turned off until $t=0$.  Then as time passes,
the wavefunction gradually renormalizes its Kruskal momentum $k$ to
$[\exp\left(k\lambda/\sqrt{M_0\pi}\right) - 1]$.  The scattering phase shift
acquired upon interaction with the incoming state will depend on how
far the state has been permitted to renormalize itself.  This can
potentially leave a trace of the time of interaction in the outgoing
wavefunction thereby carrying out some information regarding the
composition of the incoming wavepacket.  Careful scrutiny shows that
the particular feature of the self-interacted wavefunctions that could
carry this information would be the shifts $\epsilon$ and $\delta$ in
exponent on the right side of Equation~\ref{eq:iter2}.  These
quantities are expected to be shifted by the interaction with the
infaller and will not cancel in the final state like the $P_+ P_-$
shifts in the previous section.  In other words, the in-out
interactions could affect the renormalization of the Kruskal frequency
in such a way as to enable some return of information.

\section{Are Semiclassical Methods Valid?}
\label{sec:break}

The results of previous sections appear to demonstrate that
self-interactions and the in-out scattering have little effect on
the late time radiation despite dramatic effects on the individual states
involved in the computation.   In this section we will address some
consequences of the fact that the results in the previous sections also 
demonstrate that the semiclassical approximation is breaking down rapidly.   

To see this, note that the semiclassical methods are only valid when
the energies of the particles involved are small.  Now, given the
Bogliubov coefficients in Equation~\ref{eq:boginout}, an energy
eigenstate of energy $\omega$ in the region outside the black hole can
be constructed from the Kruskal states as $\phi_\omega = \int dk \,
\left[\alpha_{\omega k} \psi_k + \beta_{\omega k} \psi_k^* \right]$.
It is easily shown that this integral is dominated by modes with:
\begin{equation}
k \sim \frac{\omega}{e^{-\lambda t} 
\left(e^{\lambda r_a} - \sqrt{M_2\pi}/\lambda\right)}
\end{equation}
Since the typical late time radiation has energy $\lambda/2\pi$, this
means that the modes that contribute to the Hawking flux have a
Kruskal momentum $k$ that is growing exponentially in time.

Now, the equation $\partial S_{a0}/\partial r_a(0) = p_a(r_a(0))$
combined with Equation~\ref{eq:ps} before the collision and the
Kruskal initial condition for $S_{a0}$ gives the relation
$\frac{M_a}{\lambda} = k(e^{\lambda r_a(0)} - \sqrt{M_0 \pi}\lambda)$.
Using the solution for $\exp{\lambda r_a(0)}$ arrived at in
Section~\ref{sec:tad2part} we find that the energy of the outgoing
state prior to scattering from the incoming state is given by:
\begin{equation}
\frac{M_a}{\lambda} =
k e^{-\lambda t} \left(e^{\lambda r_a} - \frac{\sqrt{M_2
\pi}}{\lambda}\right)
+ \frac{\pi k P_+}{\lambda^2} 
\end{equation}
with $P_+ = \exp{-\lambda r_b(0)} (-\partial S_{b0}/\partial r_b(0))$.
Since $k$ is growing exponentially in time we see that the Hawking
flux emerges from states dominated by classical trajectories that have
exponentially high energies before collision with the infalling state.
At the formal level this invalidates the semiclassical approach and
suggests that the results of previous sections should not be believed.

\begin{figure}                                 
\begin{center}                                 
\leavevmode                                 
\epsfxsize=2in
\epsfysize=1in                                 
\epsffile{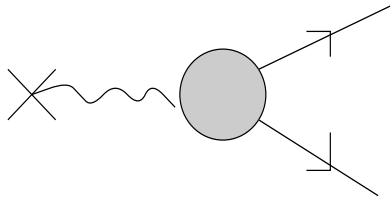}                                 
\end{center}                                 
\caption{Diagrams Contributing to The Hawking State \label{fig:create}}
\end{figure}                                 

In the calculation of the Hawking radiation, there are no real
particles in the system that have these high energies since the state
is defined to be the Kruskal vacuum.  Therefore, a common argument
says, the large shifts described above do not threaten the validity of
the semiclassical computations.  However, the calculation of the
misalignment of Kruskal and asymptotic vacua requires use of the
Bogliubov coefficients which depend on the excited states of the
system.  So, even if the system is placed in an initial state which
apparently acquires no semiclassical corrections (being a state of
Kruskal momentum zero) the calculation of Bogliubov coefficients has
limited validity because of the large corrections to the excited
states.  Put another way, the precise form of the vacuum redefinition
depends on the behaviour of modes of nonzero Kruskal momentum which
acquire large corrections.  A way to see this explicitly is in the
formula for the Hawking state $\exp(b^+\beta^*\alpha^{-1}b^+)|0>$
where $\alpha$ and $\beta$ are ``matrices'' of Bogliubov coefficients
$\alpha_{\omega k}$ and $\beta_{\omega k}$ and the operators
$b_\omega^+$ create asymptotic states.  (This can be derived as the
solution to the condition that the Hawking state is annihilated by the
Kruskal mode operators (~\cite{gn,bd}).)  This formula has the
structure of pairs of particles coming out of some loop diagram as in
Figure~\ref{fig:create}.  The matrix product $\beta^*\alpha^{-1}$
involves a sum on intermediate Kruskal states and is essentially a
summation over the particles running around the loops in
Figure~\ref{fig:create}.  As discussed earlier, the states in these
loops have energies that grow exponentially with time prior to
scattering from the incoming state.  This suggests that we simply do
not have sufficient control over the summation on virtual processes
implicit in the computation of the Hawking state to be able to believe
that the semiclassical results have exact validity.

We may fear that the results presented here are gauge dependent since
we are working with a particular parametrization of the dilaton black
hole.  However, this fear is allayed by observing that the method of
this paper is to compute the the WKB phase which is given by $\int p
\, dr$, a reparametrization invariant quantity.  So the exponentially
large scattering phase shifts that threaten the validity of the
semiclassical calculations should appear in any parametrization
including one that only uses the ``nice slices'' of~\cite{polch}.
Another objection that is sometimes raised is that arguments regarding
the breakdown of the semiclassical approximation involve
``postdiction'' whereby we try to reason about the states from which a
low energy state at the current time could have developed.  It has
been argued that difficulties with such ``postdiction'' should not be
interpreted as evidence for breakdown of the semiclassical theory
(\cite{polch}).  In fact, the problem should rather be phrased in
terms of the structure of the Hilbert space on which the field theory
is defined as we will discuss in the next section.

\subsection{Structure of The Hilbert Space and Complementarity}
As discussed above, the Hawking state is calculated by computing the
misalignment between the vacuum defined by inertial observers at the
black hole horizon and by obervers at infinity.  This misalignment
depends on the structure of the excited states of the system.  In the
presence of infalling matter, the Hilbert space contains both an
infalling and outgoing component.  In Section~\ref{sec:leading} the
states in this in-out Hilbert space were approximated by removing the
self-interactions and by having the infalling state propagate in the
static background black hole while the outgoing state propagated in
the quantum mechanical geometry thus produced.  In fact, a complete
treatment involves a study of two particle states with one infaller
and one outgoer treated with the two particle version of the
self-interacted Lagrangian~\ref{eq:lag}.  In this section we consider
such two particle states and find that the interactions between the
infaller and outgoer lead to serious difficulties with the definition
of the Hilbert space.

Using the conventions of the right-hand Figure~\ref{fig:particles} we
take $M_a/\lambda = (M_1 - M_0) /\lambda $ and $M_b/\lambda = (M_2 -
M_1)/\lambda$ to be the energies of the outgoing and incoming
particles before scattering. Similarly, take $\tilde{M}_a/\lambda =
(M_2 - \tilde{M}_1)/\lambda$ and $\tilde{M}_b/\lambda = (\tilde{M}_1 -
M_0)/\lambda$ to be the energies of the outgoing and incoming
particles after scattering.  We will begin by assuming that all these
energies are much smaller than the black hole mass so that the
semiclassical treatment is valid and show that in the kinematical
regime relevant to the Hawking radiation, we are driven well out of
the regime of validity.  The splicing prescription in
Equation~\ref{eq:bsplice} gives $\tilde{M}_b$ in terms of the incoming
energies and the position $r_c$ at which the scattering occurs.  Using
the splicing prescription and energy conservation with the two
particle equations of motion derived in Section~\ref{sec:eom}, we can
imitate the analysis of the self-interactions and the in-out
scattering in Sections~\ref{sec:selfcorr} and~\ref{sec:leading}.
Define $u_{out} = ( \exp{\lambda r_a} - \sqrt{M_2\pi}/\lambda) 
\exp{-\lambda t}$ and $v_{in} = (\exp{\lambda r_b(0)} +
\sqrt{M_2\pi}/\lambda) $.  Then, given some infalling energy
$M_b/\lambda$, it is easy to show that at late times on $\cal{I}^+$,
when $u_{out} \, v_{in} \ll M_b/\lambda^2 \ll M_0/\lambda^2$, the
following results hold in the leading order:
\begin{eqnarray}
\frac{M_a}{\lambda}  & = \omega\myprime u_{out} + 
\frac{\pi}{\lambda^2}  \frac{\omega\myprime \, M_b}{v_{in}}  &
 \\
\frac{\tilde{M}_a}{\lambda}  &= 
\omega\myprime u_{out} 
% \left[ 1  + \frac{ u_{out} v_{in}}{M_b\pi/\lambda^2} \right]
\: \: \: \: \: \: \: ; \: \: \: \: \: \: \: &
\frac{\tilde{M}_b}{\lambda}  = 
\frac{M_b}{\lambda} +
\frac{\pi}{\lambda^2}  \frac{\omega\myprime \, M_b}{v_{in}} 
%- \omega\myprime u_{out} \left[ \frac{ u_{out} v_{in}}{M_b\pi/\lambda^2}
%\right] 
\label{eq:ms} 
\end{eqnarray}
where we have defined the self-interaction renormalized frequency:
\begin{equation}
\omega\myprime = \frac{2}{\pi} \sqrt{M_0\pi}
\left(  e^{\left(k\lambda/\sqrt{M_0\pi}\right)} - 1 \right)
\end{equation}
Here we have used the fact that the energies of the particles are
assumed to be small to use the approximations made in the section on
self-interaction corrections, as well as to drop some terms of order
$O(M_b/M_0)$ which give negligible corrections to the equations of
motion of the outgoing particle prior to scattering.

\begin{figure}                                 
\begin{center}                                 
\leavevmode                                 
\epsfxsize=3in
\epsfysize=2in                                 
\epsffile{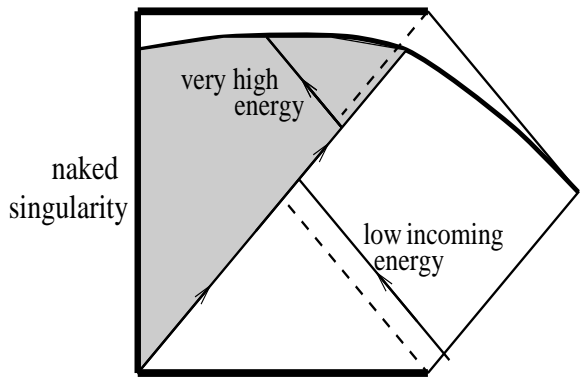}                                 
\end{center}                                 
\caption{The black hole geometry corresponding to the two-particle 
scattering process between a low energy incoming mode and a typical 
outgoing mode.  The interior geometry lies in the future of an unacceptable
naked singularity and therefore its classical interpretation is suspect. 
\label{fig:geom}}
\end{figure}

We are now in a position to extract the most important physical
consequences of the self-interacted in-out scattering.  First of all,
the results on the self-interactions and in-out scattering have told
us that the typical energy of the outgoing particle is unchanged from
the classic result and will be $\tilde{M}_a/\lambda \sim \lambda/2\pi$.
So, from Equation~\ref{eq:ms}, $\omega\myprime \sim
(\lambda/2\pi)/u_{out}$.  From the definition of $u_{out}$ we see that
$\omega\myprime$ is growing exponentially with time.  Now consider a
low energy mode dropped into the black hole with some small $M_b$.
Looking at the solution for $M_a$, and using the exponential growth of
$\omega\myprime$, we see that the state in the two particle Hilbert
space responsible for producing the late-time Hawking radiation is
dominated by classical trajectories with exponentially large energies
before scattering from the infalling state.\footnote{As discussed in the
previous section, despite the fact there are no ``real'' outgoing
particles interacting with the infaller, the computation of the vacuum
misalignment depends precisely on the behaviour of these excited states. }

The first lesson of these huge energy shifts is that states of
absurdly high energies will have to be admitted into the Hilbert space
on the early slice to produce the states on the late slice necessary
to support the Hawking radiation.  Furthermore, if we require that we
are able to observe late time Hawking radiation of reasonable
energies, then Equation~\ref{eq:ms} shows that low-energy infalling
modes will necessarily be scattered by the outgoer into states inside
the black hole with extremely large energies.

The huge energies developed by the particles can also be rewritten as
the difference in the masses of the geometries on either side of the
particle.  If we would require the mass of the internal black hole
$M_0/\lambda$ to remain fixed, the huge energies into which the
infalling particle is scattered imply that we would have to admit
states of absurdly high ADM mass into the late slice Hilbert space.
This would clearly be unphysical.  Instead, we should require that the
total energy of the system, as given by the ADM mass, remains at some
reasonable fixed values $M_2/\lambda$, and consider the states of the
system with this total energy.  In that case, the huge energies into
which the infalling particles are scattered imply that the interior
black hole geometry must have a negative mass thereby displaying a
naked singularity. (See Figure~\ref{fig:geom}.)  Both alternatives,
absurdly high energy states in the Hilbert space and naked singularity
geometries, seem quite unpalatable.

There are two complementary ways to avoid both difficulties.  The
first is to require that the radiation observed at late times has
exponentially small energies so that the states in the Hilbert space
that are responsible for producing the radiation never have energies
exceeding some finite cutoff. This allows the infalling matter to
propagate into the black hole interior without developing very high
energies on the late time slice.  This prescription therefore gives a
construction of a Hilbert space appropriate to an infalling observer.
However, this seems like an unacceptable solution to an outside
observer since it involves a drastic cutoff on the outgoing modes that
tends to zero energy at late times.  The complementary solution is to
restrict the infalling modes to have exponentially small energies.
Under this circumstance also, the states in the late time Hilbert
space remain well defined.  However, this prescription clearly imposes
severe restrictions on the physics that the infalling observer can
observe.

The dilemma seems to force the conclusion that the semiclassical
theory of Hawking radiation, even with the self-interaction
corrections, breaks down rapidly.  Furthermore, there are two
complementary, semiclassically controlled Hilbert spaces at late
times.  One is appropriate for describing the physics seen by the
infaller, and the other is appropriate to a description of the Hawking
radiation.  It is possible that these complementary Hilbert spaces
should be identified in some way giving a realization of black hole
complementarity.

\section{Conclusion}
In this paper we have studied back-reaction effects in Hawking
radiation from $1+1$ dilaton black holes.  First of all, we
constructed the effective theory produced by integrating out gravity
and the dilaton and used this to study the self-interaction of the
Hawking radiation.  We found an unusual renormalization of the Kruskal
frequency of outgoing states that nevertheless left the Hawking
temperature unchanged.  Then we studied the radiation issuing from a
dynamical background produced by a quantum mechanical state falling
into a black hole.  This calculation was carried out in an
approximation where the self-interactions were removed in order to
examine the scattering effects separately.  The in-out interaction was
found to produce large scattering phases in outgoing states that
nevertheless conspire to leave the Hawking state essentially
unchanged.  Finally, we asked whether these semiclassical conclusions
could be trusted.  We displayed the evidence that semiclassical
methods have limited validity in constructing the Hilbert space on late
slices.   We concluded from this that the structure of the 
semiclassically controlled Hilbert space supports a formulation of 
black hole complementarity.

\section{Acknowledgments}
VB would like to thank Per Kraus, Finn Larsen and Frank Wilczek for
several helpful discussions.   We would also like to thank Curt Callan for 
extensive  discussions and for collaboration in the early part of this work.
The work of VB was supported in part by DOE grant DE-FG02-91ER40671.   The 
work of HV was supported in part by the Packard Foundation, by a Pionier 
Fellowship of NWO.

\appendix
\section{Derivation of The Effective Action}
\label{sec:elim}
In this appendix we will discuss the derivation of the effective
action for dressed particles in $1+1$ dilaton gravity.  The derivation
will closely follow the work of Kraus and Wilczek (\cite{kw2}) as well
as~\cite{fmp}.   The reader is referred to these references for more
detailed discussions - the basic steps are reproduced here mainly for ease of
reference.  Figure~\ref{fig:particles} is a useful picture of the
scenario being considered. The plan of this section is as follows.
First we implement the constraints in $1+1$ dilaton gravity 
and integrate the action for an arbitrary constrained trajectory of
the geometry and the particles propagating in it. Differentiating this
action with respect to time gives a Lagrangian in which the
constraints have been incorporated.  Fixing a gauge to eliminate
redundant degrees of freedom yields the effective Lagrangian for
dressed particles.

In Section~\ref{sec:dress} we presented the Hamiltonian formulation
of $1+1$ dilaton gravity coupled to $N$ matter particles.  Since the
resulting action did not contain time derivatives of the variables
$N^t$ and $N^r$, these quantities can be integrated out generating the
constraints that $C_t = H_{tM} + H_t = 0 $ and $C_r = H_{rM} + H_r =
0 $.  By considering the linear combination of constraints
$(\pi\pi_L/2LR^2) C_r + (R\myprime/RL) C_t = 0$ where the prime
denotes $\partial/\partial r$ we find that:
\begin{equation}
\label{eq:constrt1}
-M\myprime - \sum_{i=1}^N \left[\frac{\pi\pi_L}{2LR^2} p_i 
+
 \frac{R\myprime}{LR} \sqrt{\left(\frac{p_i}{L}\right)^2 + m^2}
\right] \delta(r- r_i(t)) = 0
\end{equation}
with $M$ given by:
\begin{equation}
\label{eq:meq}
M = \frac{\pi}{4} \left(\frac{\pi_L^2}{R^2}\right) +
\frac{1}{\pi}\left[\lambda^2R^2 - \left(R\myprime/L\right)^2\right]
\end{equation}
This tells us that in the regions away from each of the particles the
quantity $M$ is constant and these constants have been labelled $M_i$
in Figure~\ref{fig:particles}. Indeed it can be shown that in the
regions between the particles the geometry is that of a black hole of mass
$M_i/\lambda$.  Since Equation~\ref{eq:constrt1} tells us that $M_i$ is
independent of $r$ we can invert the relationship in
Equation~\ref{eq:meq} to find the following expressions for $\pi_L$
and $\pi_R$:
\begin{equation}
\pi_L = \frac{2R}{\pi}\left[M_i\pi + (R\myprime/L)^2 - \lambda^2 R^2
\right]^{1/2} \: \: \: ; \: \: \pi_R =\pi_L\myprime L/ R\myprime
\end{equation}
The equation for $\pi_R$ is taken straight from the expression for the
constraint $H_{r}$ in Equation~\ref{eq:rconsrt}.   To find the
relationship between the $M_i$ we observe that the constraints at the
are consistent with $L$ and $R$ being continuous at the positions of
the particles with $\pi_L$ and $\pi_R$ free from singularities there.
Then, integrating the constraints $C_r$ and $C_t$ across the positions
of the particles gives:
\begin{eqnarray}
\pi_L(\hat{r}_i + \epsilon)  -
\pi_L(\hat{r}_i - \epsilon)  &=& \frac{ - p_i}{L(\hat{r}_i)}
 \nonumber \\
R\myprime(\hat{r}_i + \epsilon) -
R\myprime(\hat{r}_i - \epsilon) &=&
- \frac{L(\hat{r}_i)}{R(\hat{r}_i)}
\sqrt{\left(\frac{p_i}{L(\hat{r}_i)}\right)^2 + m^2}
\label{eq:piconstrt}
\end{eqnarray}
In these equations, $\hat{r}_i$ refers to the position of the
ith particle.   Simultaneous solutions of each pair of equations for
each $i$ expresses $M_i$ in terms of $M_{i-1}$ and $p_i$.

    We now follow the procedure of Kraus and Wilczek to derive the
effective action for the dressed particle trajectories.  The idea is
that the particles drag kinks in the geometry around with them and we
seek to include the contribution of these kinks to the action as part
of the gravitational dressing of the particles themselves.  To do this
we will work with a single particle in dilaton gravity - the
generalization to $N$ particles that do not cross will be trivial
since we will just have to add similar pieces for every particle in
the system.  So we consider a single particle trajectory $r_1(t)$ in
Figure~\ref{fig:particles} with $M_0/\lambda$ being the mass of the
geometry for $r < r_1$ and $M_1/\lambda$, the mass for $r > r_1$, is
the ADM mass and Hamiltonian of the system.  We begin by noting that
the action for an infinitesimal variation of the geometry and particle
trajectory is $dS = pdr_1 + \int(\pi_L \delta L + \pi_R \delta R)dr -
(M_1/\lambda)dt$. We want to integrate $S = \int dS$ for paths of the
system that obey the constraints.  The key observation is that for $r
<> r_1$, $\pi_L$ and $\pi_R$ are fixed by the constraints as a
function of $L$ and $R$.  Consequently, the Hamilton-Jacobi function
for a trajectory of the geometry, $S = \int(\pi_L \delta L + \pi_R
\delta R)$ is independent of trajectory and is a function only of
endpoints.  Let us first consider trajectories of the geometry that
leave $M_0$ and $M_1$ fixed.  In the regions $r<r_1$ and $r>r_1$ we
can follow~\cite{fmp} to integrate $S$ from a configuration A to a
configuration B as follows.  Starting from any configuration we can
integrate $L(r)$ along a path of constant $R(r)$ to a configuration
with $(R\myprime/L) = \sqrt{M\pi - \lambda^2 R^2}$.  This
configuration has $\pi_L = \pi_R = 0$ and is therefore static.  Then
holding this relation between $R\myprime$ and $L$ fixed we integrate
to some other standard static configuration.  The second leg of the
integration has $\pi_L = \pi_R = 0$ and so does not contribute
anything.  To integrate between any pair of configurations $A$ and $B$
we integrate in this manner from $A$ to the standard configuration and
from there to $B$.  This gives the following action for the motion of
the geometry in the regions $r > r_1$ and $r < r_1$ in which we have
dropped the constant arising from the lower limit of the $L$ integration:
\begin{equation}
S = \int_0^{r_1 - \epsilon} dr\, F_0(r,t) + \int_{r_1 +
\epsilon}^\infty dr\, F_1(r,t)
\label{eq:tempac}
\end{equation}
where we define $B_i = \sqrt{\lambda^2 R(r)^2 - M_i\pi}$ and 
\begin{equation}
F_i = \frac{2 R L}{\pi} \sqrt{\left(\frac{R\myprime}{L}\right)^2 -
B_i^2}
+ \frac{2 R R\myprime}{\pi} \ln\left[\frac{(R\myprime/L) - 
\sqrt{(R\myprime/L)^2 - B_i^2}}{B_i} \right]
\end{equation}
Equation~\ref{eq:tempac} is the action for a trajectory for which
there are no variations of the geometry at $r_1$, the position of the
shell and which keeps the mass $M_1$ constant.  To find the action for
a general trajectory of the geometry consider a general variation of
Equation~\ref{eq:tempac} with respect to $L$ and $R$.  We find that:
\begin{eqnarray}
dS = &\int_0^\infty dr\left[\pi_L \delta L + \pi_R \delta_R \right]
   + \left[\frac{\partial F_1}{\partial R\myprime}(r_1 + \epsilon)
     - \frac{\partial F_1}{\partial R\myprime}(r_1 - \epsilon) \right]
      d\hat{R} + \nonumber \\
&   \int_{r_1}^{\infty} dr \, \frac{\partial F_1}{\partial M_1} \, dM_1 
\label{eq:var}
\end{eqnarray}
where $d\hat{R}$ represents the total variation in $R$ at the position
of the particle and $dM_1$ is the variation in the ADM mass caused by
an arbitrary variation of $L$ and $R$.  The term proportional to
$d\hat{R}$ arises because $R\myprime$ is constrained to be
discontinuous across the position of the particle
(Equation~\ref{eq:piconstrt}) and represents the contribution of the
kink in the geometry.  The terms proportional to $dM_1$ arises because
a general variation of the geometry can change the mass of the system.
There is no contribution from $dM_0$ because we assume that the mass
of the background black hole is held fixed.  Since we must require
that $\delta S/\delta R = \pi_R$ and $\delta S/\delta L = \pi_L$ we
must subtract the integrated contribution of the two anomalous
variations on the right of Equation~\ref{eq:var}.  Putting everything
together we arrive at the following action for a constrained variation
of dilaton-gravity coupled to a single particle:
\begin{eqnarray}
S = &\int_0^{r_1 - \epsilon} dr \, F_0 + \int_{r_1 + \epsilon}^\infty
dr \, F_1  \: -  \: 
\int_0^t dt\, \frac{d\hat{R}}{dt} 
    \left[\frac{\partial F_1}{\partial R\myprime}(r_1 + \epsilon)
     - \frac{\partial F_1}{\partial R\myprime}(r_1 - \epsilon) \right]
- \nonumber \\
&\int_0^t dt\, \int_{r_1}^{\infty} dr \frac{dM_1}{dt} 
                       \frac{\partial F_1}{\partial M_1} 
\: - \:  \int_0^t dt\, \frac{M_1}{\lambda}
\label{eq:totac}
\end{eqnarray}
Here $\hat{R} = R(r_1)$ is the value of $R$ at the position of the
particle.  Although we have not explicitly considered the contribution
of a variation of $r_1$ to the action, the integrability of the
equations ensures that such a piece has been included as can be
checked by explicitly differentiating back.

We are now in a position to derive the effective constrained
Lagrangian for this system by differentiating the action in
Equation~\ref{eq:totac}.  Indeed, writing $L= dS/dt$, and using the
integrated constraints for a massless particle ($m=0$) in
Equation~\ref{eq:piconstrt} in imitation of Kraus and Wilczek
(~\cite{kw2}), we find the following Lagrangian:
\begin{eqnarray}
L = & \dot{r}_1 \frac{2 \hat{R}\hat{L}}{\pi} \left[
\left. \sqrt{\left( R\myprime/ L \right)^2 - B_0^2 }\right|_< -
\left. \sqrt{\left( R\myprime/L \right)^2 - B_1^2 }\right|_> \right]
- \nonumber \\
&
\eta \frac{2 \hat{R} \dot{\hat{R}} }{\pi}
\ln\left[
\frac{(R\myprime/L)_> - \eta
\left.\sqrt{\left( R\myprime/L \right)^2 - B_1^2 }\right|_>}
{(R\myprime/L)_< - \eta
\left.\sqrt{\left( R\myprime/L \right)^2 - B_0^2 }\right|_<}
\right] + \nonumber \\
&
\int_0^{r_1 - \epsilon} dr \, \left[\pi_L \dot{L} + \pi_R \dot{R} \right]
+ 
\int_{r+1 + \epsilon}^{\infty} dr \, \left[\pi_L \dot{L} + \pi_R
\dot{R} \right] 
-  \frac{M_1}{\lambda}
\end{eqnarray}
In this equation $>$ and $<$ indicate positions on either side of the
particle at $r_1$ that are infinitesimally close by, but not subject
to the constraint in Equation~\ref{eq:piconstrt}. Furthermore, $\eta =
\pm = sign(p)$.  The expression on the second line is the contribution
of a kink in the geometry to the effective Lagrangian and all the
effort of integrating up the action and differentiating back has been
designed to correctly pick up this contribution.  We are now free to
fix a gauge for $L$ and $R$.  A particularly convenient gauge is $L=1$
and $R = \exp{\lambda r}$ which yields the following effective
Lagrangian:
\begin{equation}
L = \left(\frac{2}{\pi}\right)
 \dot{r} \left\{ e^{\lambda r} \left( \sqrt{M_{0}\pi} -
\sqrt{M_1 \pi} \right) 
- \eta \lambda e^{2\lambda r}
\ln\left(\frac{e^{\lambda r} - \eta \sqrt{M_1
\pi}/\lambda}{e^{\lambda r} - \eta \sqrt{M_{0} \pi}/\lambda}
\right) \right\} \: - \: \frac{M_1}{\lambda} 
\end{equation}
This equation is to be interpreted as $L = p_{c1} \dot{r_1} - H$ where
$H = M_1 / \lambda$ is the Hamiltonian of the system and $p_{c1}$ is
the effective canonical momentum.  In fact, it is convenient to
subtract out the constant contribution of the background black hole
to the Hamiltonian to write the Hamiltonian as $H = (M_1 -
M_0)/\lambda$.  To get the effective Lagrangian for multiple particles
we need only add a similar $p \dot{r}$ term for each particle and pick
the new ADM mass as the Hamiltonian.  This has been done is
Section~\ref{sec:dress}.

\section{Integrability Of The Dressed Mechanics}
\label{sec:integ}
In this appendix we will demonstrate that the quantities $M_i$ that
determine the masses of the geometries in between the particles form a
system of $N$ conserved quantities with mutually vanishing Poisson
brackets.  This shows that the effective dynamics of the system is
classically integrable so long as the particles do not cross each
other.   To show that $dM_i/dt = 0$ we start with the observation that
$dM_N/dt = 0$ since $M_N/\lambda$ is the global Hamiltonian of the
system.     Next, observe that Equation~\ref{eq:can} defining
$p_{ci}$ can be written as a definition of $M_{i-1}$ in terms of the
dynamical of variables of the ith particle and $M_i$: $M_{i-1} =
M_{i-1}(p_i,r_i,Mi)$.   Let us take as the induction hypothesis that
$dM_{i+1}/dt = 0$.   We can write:
\begin{equation}
\frac{dM_i}{dt} = 
\frac{\partial M_i}{\partial M_{i+1}} 
%\right|_{p_{c(i+1)},r_{i+1}}
\frac{dM_{i+1}}{dt}
+
\frac{\partial M_i}{\partial p_{c(i+1)}} 
%\right|_{M_{i+1},r_{i+1}}
\frac{dp_{c(i+1)}}{dt}
+
\frac{\partial M_i}{\partial r_{i+1}} 
%\right|_{p_{c(i+1)},M_{i+1}}
\frac{dr_{i+1}}{dt}
\label{eq:dMdt}
\end{equation}
We now compute the various partial and total derivatives required to
evaluate this expression and plug back in.   First of all note that
Equation~\ref{eq:can} for the canonical momenta can be written in the form
$p_{c(i+1)} = f(M_i, r_{i+1}) - f(M_{i+1},r_{i+1})$ for a suitably
chosen function $f$.   Define the quantities:
\begin{eqnarray}
A(M_i,r_{i+1}) &=& \frac{ \partial f(M_i,r_{i+1})}{\partial M_i}
\nonumber \\
B(M_i,M_{i+1},r_{i+1}) &=& \frac{\partial f(M_i,r_{i+1})}{\partial r_{i+1}}
- \frac{\partial f(M_{i+1},r_{i+1})}{\partial r_{i+1}}
\end{eqnarray}
We can now differentiate both sides of the equation for $p_{c(i+1)}$
to get:
\begin{eqnarray}
1 &=& A(M_i,r_{i+1}) \, \frac{\partial M_i}{\partial p_{c(i+1)}} \nonumber \\
0 &=& B(M_i,M_{i+1},r_{i+1}) + 
A(M_i,r_{i+1}) \frac{\partial M_i}{\partial r_{i+1}}
\end{eqnarray}
These expressions can be solved to find the partial derivatives in
Equation~\ref{eq:dMdt}.   Finally, we want to compute $dr_{i+1}/dt$
and $dp_{c(i+1)}/dt$.   Since $M_N/\lambda$ is the Hamiltonian these
are given by $dr_{i+1}/dt  = (1/\lambda) \partial M_N/\partial p_{i+1}$ and 
$dp_{c(i+1)}/dt = (-1/\lambda) \partial M_N/\partial r_{i+1}$ where
the partial derivatives are taken while holding all the other
canonical pairs $\{r_i,p_{ci}\}$ constant.   We can use the chain of
definitions of $M_i$ in terms of $M_{i-1}$ to compute these
derivatives as:
\begin{equation}
\frac{d r_{i+1}}{dt} = \frac{K}{\lambda} \frac{\partial M_{i+1}}{\partial p_{c(i+1)}} 
\: \: ; \: \:
\frac{d p_{c(i+1)}}{dt} = \frac{- K}{\lambda} \frac{\partial M_{i+1}}{\partial r_{i+1}} 
\end{equation}
with $K$ defined as:
\begin{equation}
K \equiv \left( \frac{\partial M_N}{\partial M_{N-1}}
\right)_{r_N,p_{cN}} 
\cdots
\left( \frac{\partial M_{i+2}}{\partial M_{i+1}}
\right)_{r_{i+2},p_{c(i+2)}}
\end{equation}
Putting these expressions back into Equation~\ref{eq:dMdt} we find that
\begin{eqnarray}
\frac{d M_i}{dt} = & \frac{\partial M_i}{\partial M_{i+1}} \frac{d
M_{i+1}}{dt} + 
\frac{-1}{A(M_i,r_{i+1})} 
\frac{K B(M_i,M_{i+1},r_{i+1})}{A(M_{i+1},r_{i+1})}
+ \nonumber \\
&
\frac{- B(M_i,M_{i+1},r_{i+1})}{A(M_i,r_{i+1})} 
\frac{- K}{A(M_{i+1},r_{i+1})} = 0
\end{eqnarray}
where we have used the induction hypothesis that $dM_{i+1}/dt = 0$
that is satisfied for $M_N$. 

Having shown that the $M_i$ are conserved we show that they
have mutually vanishing Poisson brackets.  The Possion brackets are
given by: $\{M_i,M_j\} = \sum_k (\partial M_i/\partial r_k \, \partial
M_j/\partial p_k - \partial M_j/\partial p_k \, \partial M_i/\partial
r_k)$ where each partial derivative is evaluated while holding all
other canonical variables constant.   We can use the
Equations~\ref{eq:can} for the canonical momenta as an implicit chain
of definitions of $M_i$ in terms of $M_{i-1}$: $M_i = M_i(p_{ci}, r_i,
M_{i-1})$.   This permits us to write $M_i = M_i(\{p_{cj}, r_j\}_{j\leq
i})$.   So consider $\{M_i,M_j\}$ for $j > i$.   Because each $M_i$ is
defined in terms of $p_j$ and $r_j$ for $j \leq i$, this Possion
bracket is given by:
$\{M_i,M_j\} = \sum_{k\leq i} (\partial M_i/\partial r_k \, \partial
M_j/\partial p_k - \partial M_j/\partial p_k \, \partial M_i/\partial
r_k)$.   Now define $G(i,k) = (\partial M_i/\partial M_{i-1}) \cdots
(\partial M_{k+1}/\partial M_k)$ with $G(i,i) \equiv 1$.   Then
$\partial M_i/\partial r_k = G(i,k) \partial M_k/\partial r_k$ and
$\partial M_i/\partial p_{ck} = G(i,k) \partial M_k/\partial p_{ck}$.
This gives us the result that:
\begin{equation}
\{M_i,M_j\} = \sum_{k\leq i} G(i,k)G(j,k) \left[ \frac{\partial M_k}{\partial
r_k} \, \frac{\partial M_k}{\partial p_{ck}}  -
\frac{\partial M_k}{\partial
p_{ck}} \, \frac{\partial M_k}{\partial r_{k}}
\right] = 0
\end{equation}
This proves that the Possion brackets of the $M_i$ vanish.

We have seen that the $M_i$ form a set of $N$ mutually commuting
quantities with vanishing Possion brackets.   This tells us that the
system of $N$ gravitationally dressed particles is classically
integrable so long as the particles do not cross.

\section{Computation of One Particle Dressed Wavefunctions}
\label{sec:onewave}
In this appendix we will solve the WKB self-consistency conditions in
Equation~\ref{eq:self} in order to compute the one particle dressed
wavefunctions.   It is useful to define $\epsilon$ and $\delta$ as the
following small quantities:
\begin{equation}
\frac{\sqrt{M_1 \pi}}{\lambda} =\frac{\sqrt{M_0 \pi}}{\lambda} +
\epsilon
\: \: \: \: ; \: \: \: \:
e^{\lambda r_0} = \frac{\sqrt{M_1 \pi}}{\lambda} + \delta =
\frac{\sqrt{M_0 \pi}}{\lambda} + \epsilon + \delta
\end{equation}
In these equations $\epsilon$ measures the difference between the
boundary of the trapped surfaces for self-interacting particles and
the event horizon associated with the background black hole.  The
small quantity $\delta$ measures how far the initial position of the
particle deviates from the boundary of the trapped surfaces.  Using
the initial condition $S_0$ and the equation of
motion~(\ref{eq:1motion}) the self-consistency
conditions~(\ref{eq:self}) can now be written as:
\begin{eqnarray}
\delta &=& e^{-\lambda t} \left( e^{\lambda r} - \frac{\sqrt{M_0
\pi}}{\lambda} - \epsilon \right) \label{eq:iter1} \\ 
\epsilon &=&
\delta \left[ \exp\left(\frac{k + \epsilon}{\sqrt{M_0 \pi}/\lambda +
\delta + \epsilon}\right) - 1 \right]
\label{eq:iter2}
\end{eqnarray}
In order to solve these equations we have to specify the kinematical
regimes that are of interest to us so that we can make suitable
approximations.  We are interested in observing the radiation at late
times on ${\cal I}^+$.  So we will take $e^{\lambda r} \gg \sqrt{M_0
\pi}/\lambda$ so that the observations are made far from the horizon
of the black hole while $u = e^{-\lambda t} (e^{\lambda r} - \sqrt{M_0
\pi}/\lambda)$ is small.  We will also assume that the outgoing state
has an energy much smaller than $M_0/\lambda$, the energy of the black
hole, because the semiclassical methods are not reliable for extremely
energetic states.  This choice of kinematical regime implies that both
$\epsilon$ and $\delta$ are much smaller that $\sqrt{M_0
\pi}/\lambda$.  It is shown in Section~\ref{sec:selfcorr} that Hawking
radiation observed at a time $t$ arises from states with Kruskal
momenta that grow rapidly with time and this tells us that for the
physics questions of interest to us we can drop the $\epsilon$ and the
$\delta$ in the exponent of Equation~\ref{eq:iter2} to leading order.
Having linearized the equations in this way we solve them to find
that:
\begin{eqnarray}
\label{eq:soln}
\delta & = \frac{e^{-\lambda t} (e^{\lambda r} - \sqrt{M_0
\pi}/\lambda)}{1 + e^{-\lambda t} (\gamma - 1) }
\: \: \: ; \: \: \:
\epsilon = \frac{e^{-\lambda t} (e^{\lambda r} - \sqrt{M_0
\pi}/\lambda)}{1 + e^{-\lambda t} (\gamma - 1) } 
(\gamma - 1) \\
& \gamma \equiv \exp\left( \frac{k\lambda}{\sqrt{M_0 \pi}} +
f(r,t,k,M_0) \right)
\end{eqnarray}
In the equation for $\gamma$, the function $f$ is zero to leading
order - we have included it merely to remind us that there are
additional subleading dependencies that we have omitted.

     We are now ready to calculate the phase of the dressed
wavefunction $S(r,t) = S_0(r_0) + R(r) - R(r_0) - (M_1 - M_0)t/\lambda$
where $R = \int p_c \, dr$ is given in Equation~\ref{eq:R}.  Since we
have only calculated $\epsilon$ and $\delta$ to leading order we
should linearize $R$ in terms of these small quantities.  Keeping
terms of order $\epsilon$, $\delta$, $\epsilon\ln{\epsilon}$ and
$\delta\ln{\delta}$, we substitute our solutions for these quantities
to find:
% Doing this, the leading order expression for the phase
%of the dressed wavefunction is:
%\begin{eqnarray}
%S(r,t) &=& \frac{2k}{\pi}(\epsilon + \delta) + 
%\frac{2}{\pi} \frac{\sqrt{M_0\pi}}{\lambda} \times
%\nonumber \\
%& &  \left[ \epsilon \left(
%1 + \ln\left(e^{\lambda r} - \frac{\sqrt{M_0
%\pi}}{\lambda} \right) - \lambda t\right) 
%+ \delta\ln{\delta}
%- (\delta + \epsilon) \ln(\delta + \epsilon) \right]
%\end{eqnarray}
%Substituting our solution for $\epsilon$ and $\delta$ into this
%equation gives:
\begin{eqnarray}
S(r,t) = &\frac{2}{\pi}
\frac{\sqrt{M_0 \pi}}{\lambda} e^{-\lambda t} \left(e^{\lambda r} -
\frac{\sqrt{M_0 \pi}}{\lambda} \right) (\gamma -1) \times \nonumber \\
&\frac{
\left[ 1+ f(r,t,k,M_0) + \ln\left(1 + e^{-\lambda t}(\gamma -1
)\right)  \right] }
{1 + e^{-\lambda t}(\gamma -1)}
\end{eqnarray}
As discussed earlier we expect $f$ to be small and from
Equation~\ref{eq:soln} we see that if $\epsilon$ is
assumed to be small then at late times on $\cal{I}^+$ (large $r$ with
$e^{-\lambda t}(e^{\lambda r} - \sqrt{M_0\pi}/\lambda)$ small)
$e^{-\lambda t}(\gamma - 1)$ must be small also.    This yields the
leading order dressed wavefunction in Equation~\ref{eq:leading}


\begin{thebibliography}{10}

\bibitem{hawking1}
S.~W. Hawking.
\newblock Particle creation by black holes.
\newblock {\em Commun. Math. Phys.}, 43:199--220, 1975.

\bibitem{hawking2}
S.~W. Hawking.
\newblock Breakdown of predictability in gravitational collapse.
\newblock {\em Phys. Rev. D}, 14(10):2460--2473, November 1976.

\bibitem{thooft1}
G.'t Hooft.
\newblock The black hole interpretation of string theory.
\newblock {\em Nucl. Phys. B}, 335:138--154, 1990.

\bibitem{stu}
L.~Susskind, L.~Thorlacius, and J.~Uglum.
\newblock The stretched horizon and black hole complementarity.
\newblock {\em Phys. Rev. D}, 48(8):3743--3761, October 1993.

\bibitem{sv21}
K.~Schoutens, H.~Verlinde, and E.~Verlinde.
\newblock Quantum black hole evaporation.
\newblock {\em Phys. Rev. D}, 48(6):2670--2785, September 1993.

\bibitem{sv22}
K.~Schoutens, H.~Verlinde, and E.~Verlinde.
\newblock Black hole evaporation and quantum gravity.
\newblock {\em CERN-TH.7142/94}, January 1994.

\bibitem{kv2}
Y.~Kiem, H.~Verlinde, and E.~Verlinde.
\newblock Quantum horizons and complementarity.
\newblock {\em CERN-TH.7469/95, PUPT-1504}, January 1995.

\bibitem{lowe}
D.A. Lowe, J.~Polchinski, L.~Susskind, L.~Thorlacius, and J.~Uglum.
\newblock Black hole complementarity versus locality.
\newblock {\em NSF-ITP-95-47, hep-th/9506138}, June 1995.

\bibitem{keski}
E.~Keski-Vakkuri, G.~Lifschytz, S.~Mathur, and M.~Ortiz.
\newblock Breakdown of the semi-classical approximation at the black hole
  horizon.
\newblock {\em MIT-CTP-2341, hep-th/9408039}, August 1994.

\bibitem{kw2}
P.~Kraus and F.~Wilczek.
\newblock Self-interaction correction to black hole radiance.
\newblock {\em PUPT 1490, IASSNS 94/61}, August 1994.

\bibitem{kw3}
P.~Kraus and F.~Wilczek.
\newblock Effect of self-interaction on charged black hole radiance.
\newblock {\em PUPT 1511, IASSNS 94/101}, November 1994.

\bibitem{fmp}
W.~Fischler, D.~Morgan, and J.~Polchinski.
\newblock Quantization of false-vacuum bubbles: A hamiltonian treatment of
  gravitational tunneling.
\newblock {\em Phys. Rev. D}, 42(12):4042--4055, December 1990.

\bibitem{cghs}
C.C. Callan, S.B. Giddings, J.A. Harvey, and A.~Strominger.
\newblock Evanescent black holes.
\newblock {\em Phys. Rev. D}, 45(4):R1005--R1010, February 1992.

\bibitem{kw1}
P.~Kraus and F.~Wilczek.
\newblock Some applications of a simple stationary line line lement for the
  schwarschild geometry.
\newblock {\em PUPT 1474}, June 1994.

\bibitem{witten}
E.~Witten.
\newblock String theory and black holes.
\newblock {\em Phys. Rev. D}, 44(2):314--324, July 1991.

\bibitem{mandal}
G.~Mandal, A.M. Sengupta, and S.R. Wadia.
\newblock Classical solutions of two-dimensional string theory.
\newblock {\em Mod. Phys. Lett. A}, 6:1685--1692, March 1991.

\bibitem{rst}
J.G. Russo, L.~Susskind, and L.~Thorlacius.
\newblock The endpoint of hawking radiation.
\newblock {\em Phys. Rev. D}, 46(8):3444--3449, October 1992.

\bibitem{thorlacius}
L.~Thorlacius.
\newblock Black hole evolution.
\newblock {\em hep-th/9411020}, November 1994.

\bibitem{gn}
S.~B. Giddings and W.~M. Nelson.
\newblock Quantum emission from two-dimensional black holes.
\newblock {\em UCSBTH-92-15, hep-th/9204072}, April 1992.

\bibitem{bd}
N.D. Birrell and P.C.W. Davies.
\newblock {\em Quantum Fields in Curved Space}.
\newblock Cambridge University Press, 1982.

\bibitem{gutz}
M.~Gutzwiller.
\newblock {\em Chaos in Classical and Quantum Mechanics}.
\newblock Springer-Verlag, 1990.

\bibitem{peet}
A.~Peet, L.~Susskind, and L.~Thorlacius.
\newblock Information loss and anomalous scattering.
\newblock {\em Phys. Rev. D}, 46(8):3435--3443, October 1992.

\bibitem{polch}
J.~Polchinski.
\newblock String theory and black hole complementarity.
\newblock {\em hep-th/9507094, NSF-ITP-95-63}, July 1995.

\end{thebibliography}
\end{document}